\documentclass[aps,showpacs,nofootinbib,superscriptaddress,preprintnumbers]{revtex4}

\usepackage{graphicx}
\usepackage{graphics}
\usepackage{amssymb}
\usepackage{amsmath}
\usepackage{bm}

\newcommand{\g}{\gamma}

\newcommand{\open}{\sphericalangle}
\newcommand{\sT}{\scriptstyle T}

\begin{document}
\allowdisplaybreaks[2]

%%%%%%%%%%%%%%%%%%%%%%%%%%%%%%%%%%%%%%%%%%%%%%%%%%%%%%%%%%%%%%%%%%%%%%%%
\title{
Asymmetries involving dihadron fragmentation functions: from DIS to $e^+e^-$ 
annihilation}
%%%%%%%%%%%%%%%%%%%%%%%%%%%%%%%%%%%%%%%%%%%%%%%%%%%%%%%%%%%%%%%%%%%%%%%%

\author{Alessandro Bacchetta}
\email{alessandro.bacchetta@jlab.org}
\affiliation{Theory Center, Jefferson Lab, 12000 Jefferson Ave, Newport News,
  VA 23606, USA}

\author{Federico Alberto Ceccopieri}
\email{federicoalberto.ceccopieri@fis.unipr.it}
\affiliation{Dipartimento di Fisica, Universit\`a di Parma, \\
and INFN, Gruppo Collegato di Parma, I-43100 Parma, Italy}

\author{Asmita Mukherjee}
\email{asmita@phy.iitb.ac.in}
\affiliation{Physics Department, Indian Institute of Technology Bombay,
Powai, Mumbai 400076, India}

\author{Marco Radici}
\email{marco.radici@pv.infn.it}
\affiliation{Istituto Nazionale di Fisica Nucleare, Sezione di Pavia, I-27100 
Pavia, Italy}

%%%%%%%%%%%%%%%%%%%%%%%%%%%%%%%%%%%%%%%%%%%%%%%%%%%%%%%%%%%%
\begin{abstract}
Using a model calculation of dihadron fragmentation functions, we fit the spin
asymmetry recently extracted by HERMES for the semi-inclusive pion pair 
production in deep-inelastic scattering on a transversely polarized proton
target. By evolving the obtained dihadron fragmentation functions, we make
predictions for the correlation of the angular distributions of two pion pairs
produced in electron-positron annihilations at BELLE kinematics. 
Our study shows 
that the combination of two-hadron inclusive deep-inelastic scattering and 
electron-positron annihilation measurements can provide a valid alternative to 
Collins effect for the extraction of the quark transversity distribution in the 
nucleon. 
\end{abstract}

%%%%%%%%%%%%%%%%%%%%%%%%%%%%%%%%%%%%%%%%%%%%%%%%%%%%%%%%%%%%%%%%%%%%%%%%


\preprint{JLAB-THY-08-917}

%\date{\today}

\maketitle

%%%%%%%%%%%%%%%%%%%%%%%%%%%%%%%%%%%%%%%%%%%%%%%%%%%%%%%%%%%%
\section{Introduction}
\label{sec:intro}

Dihadron fragmentation functions (DiFF) describe the hadronization of a quark
in two hadrons plus other unobserved fragments. In their simplest form, they
represent  the probability that at some hard scale a parton hadronizes in two 
hadrons with fractional energies $z_1$ and $z_2$. They were introduced for the 
first time when studying the $e^+ e^- \rightarrow h_1 h_2 X$ process in the 
context of jet calculus~\cite{Konishi:1978yx}. They are in fact necessary to 
guarantee the factorization of all collinear singularities for such a process 
at next-to-leading order (NLO) in the strong coupling 
constant~\cite{deFlorian:2003cg}.

In experiments, often not only the fractional energies of the two hadrons are 
measured, but also their invariant mass $M_h$ (see, e.g., 
Refs.~\cite{Acton:1992sa,Abreu:1992xx,Buskulic:1995gm}). Hence, it is useful 
to introduce extended DiFF (in analogy with extended fracture 
functions~\cite{Grazzini:1997ih}), which are explicitly depending on $M_h$. 
Their definition and properties have been analyzed up to subleading 
twist~\cite{Bianconi:1999cd,Bacchetta:2003vn}. Their evolution equations are 
known~\cite{Ceccopieri:2007ip} and presently solved in the leading logarithm 
approximation (LL), and there are valid arguments to assume that they can be 
factorized and are universal, similarly to what happens for extended fracture 
functions~\cite{Grazzini:1997ih}. In fact, (extended) DiFF can appear also in 
two-particle-inclusive deep-inelastic scattering (SIDIS) and in hadron-hadron 
collisions. 

DiFF can be used as analyzers of the polarization state of the fragmenting 
parton~\cite{Efremov:1992pe,Collins:1994kq,Artru:1996zu}. Because of this, they 
have been proposed as tools to investigate the spin structure of the nucleon, 
in particular to measure the transversity distribution $h_1^q$ of a parton $q$ 
in the nucleon $N$ (see Ref.~\cite{Barone:2003fy} for a review). The $h_1^q$, 
together with the momentum $f_1^q$ and helicity $g_1^q$ distributions, fully 
characterizes the (leading-order) momentum/spin status of $q$ inside $N$, if
quark transverse momentum is integrated over. Transversity is a chiral-odd
function and  needs to appear in a cross section accompanied by another 
chiral-odd function. The simplest example is the fully transversely polarized 
Drell-Yan process, where $h_1^q$ appears multiplied by its antiquark partner 
$\overline{h}_1^q$~\cite{Ralston:1979ys}. Although this process is
theoretically very clean, it appears to be experimentally very
challenging~\cite{Martin:1999mg}, at least at present facilities (the same 
finding is confirmed for proton-proton collisions leading to prompt 
photon~\cite{Mukherjee:2003pf} and semi-inclusive pion 
production~\cite{Mukherjee:2005rw}).

An alternative approach, so far the only fruitful one, is to turn to SIDIS and
measure the correlation between the transverse polarization of the target and 
the transverse momentum of the final hadron, which involves a convolution of 
$h_1^q$ with the chiral-odd Collins fragmentation function 
$H_1^{\perp\,q}$~\cite{Collins:1993kk}. The resulting asymmetry has already 
been measured at HERMES~\cite{Airapetian:2004tw,Diefenthaler:2007}, and at 
COMPASS~\cite{Alexakhin:2005iw,Ageev:2006da,Levorato:2008tv}. The knowledge of 
the Collins function is required to extract the transversity distribution. This 
can be obtained through the measurement of azimuthal asymmetries in 
$e^+ e^- \rightarrow \pi^+ \pi^- X$ with almost back-to-back 
pions~\cite{Boer:1997mf}. The BELLE collaboration at KEK has measured this 
asymmetry~\cite{Seidl:2008xc}, making the first-ever extraction of $h_1^q$ 
possible from the global analysis of SIDIS and $e^+ e^-$ 
data~\cite{Anselmino:2007fs}. 

At present, large uncertainties still affect this analysis and the resulting 
parametrization of $h_1^q$. The most crucial issue is the treatment of 
evolution effects, since the BELLE and the HERMES/COMPASS measurements happened 
at two very different scales: $Q^2\sim 100$ and $\langle Q^2 \rangle = 2.5$ 
GeV$^2$, respectively. Both $h_1\otimes H_1^\perp$ and 
$H_1^\perp \otimes \overline{H}_1^\perp$ convolutions involve 
transverse-momentum dependent functions (TMD)~\cite{Boer:1997nt,Boer:2008fr} 
whose behaviour upon scale change should be described in the context of 
Collins-Soper factorization~\cite{Collins:1981uk,Ji:2004wu} (see also 
Refs.~\cite{Ceccopieri:2005zz,Bacchetta:2007wc}). However, the global analysis 
of Ref.~\cite{Anselmino:2007fs} neglects any change of the partonic transverse 
momentum with the scale $Q^2$ leading to a possible overestimation of 
$h_1$~\cite{Boer:2001he,Boer:2008fr,Boer:2008mz}. It would be desirable, 
therefore, to have an independent way to extract transversity, involving 
collinear fragmentation functions. Here, we consider the semi-inclusive 
production of two hadrons inside the same current jet. 

As already explained, the fragmentation $q\rightarrow (\pi^+ \pi^-) X$ is
described by an (extended) DiFF. When the quark is transversely polarized, 
$q^\uparrow$, a correlation can exist between its transverse polarization 
vector and the normal to the plane containing the two pion momenta. This effect 
is encoded in the chiral-odd polarized DiFF $H_1^{\open\,q}$ via the dependence 
on the transverse component of the pion pair relative momentum 
$R_{\sT}$~\cite{Bianconi:1999cd}. The function $H_1^{\open\,q}$ can be 
interpreted as arising from the interference of $(\pi^+ \pi^-)$ being in two
states with different angular 
momenta~\cite{Collins:1994ax,Jaffe:1998hf,Radici:2001na,Bacchetta:2006un}. 
Since the transverse momentum of the hard parton is integrated out, the cross 
section can be studied in the context of collinear factorization and its 
polarized part contains the factorized product 
$h_1^q H_1^{\open\,q}$~\cite{Radici:2001na,Bacchetta:2002ux}. The HERMES 
collaboration has recently measured such spin asymmetry using transversely 
polarized proton targets~\cite{Airapetian:2008sk}; the COMPASS collaboration 
performed the same measurement on a deuteron target~\cite{Martin:2007au} and 
should soon release data using a proton target. 

Similarly to the Collins effect, the unknown $H_1^{\open\,q}$ has to be
extracted from electron-positron annihilation, specifically by measuring the 
angular correlation of planes containing two pion pairs in the 
$e^+ e^- \rightarrow (\pi^+ \pi^-)_{\rm jet1}\, (\pi^+ \pi^-)_{\rm jet2} X$ 
process~\cite{Artru:1996zu,Boer:2003ya}. The BELLE collaboration is analyzing 
data for this angular correlation~\cite{Abe:2005zx,Hasuko:2003ay}. Therefore, 
it seems timely to use available models for extended DiFF to make predictions 
for the $e^+ e^-$ azimuthal asymmetry at BELLE kinematics. Since evolution 
equations for extended DiFF are available at NLO~\cite{Ceccopieri:2007ip}, at 
variance with the Collins effect the asymmetries with inclusive hadron pairs in 
SIDIS and $e^+ e^-$ can be correctly connected when the scale is ranging over 
two orders or magnitude. Therefore, the option of using the semi-inclusive 
production of hadron pairs inside the same jet seems a theoretically clean way 
to extract transversity~\cite{Boer:2008mz}. Finally, we point out that pair 
production in polarized hadron--hadron collisions allows in principle to 
"self-sufficiently" determine all the unknown DiFF and 
$h_1$~\cite{Bacchetta:2004it}. The STAR collaboration has recently presented 
data on this kind of measurement~\cite{Yang:2008}.
%Ruizhe Yang, Probing transversity with interference fragmentation function 
%in pp collision at $\sqrt{s}$ = 200 GeV  
%PKU-RBRC Workshop on Transverse Spin Physics,
%Beijing, June 30th-July 4th, 2008
%http://rchep.pku.edu.cn/workshop/0806/20080627-iff.pdf

The paper is organized as follows. In Sec.~\ref{sec:fit}, using the model 
calculation of DiFF from Ref.~\cite{Bacchetta:2006un}, we fit the spin 
asymmetry recently extracted by HERMES for the SIDIS production of 
$(\pi^+ \pi^-)$ pairs on transversely polarized 
protons~\cite{Airapetian:2008sk}. In Sec.~\ref{sec:evolution}, we describe how
we calculate the evolution of the involved extended DiFF starting from the 
HERMES scale up to the BELLE scale. In Sec.~\ref{sec:predictions}, we 
illustrate the predictions for the correlation of angular distributions of two 
pion pairs produced in $e^+ e^-$ annihilations at BELLE kinematics. Finally, in 
Sec.~\ref{sec:conclusions} we draw some conclusions.

%%%%%%%%%%%%%%%%%%%%%%%%%%%%%%%%%%%%%%%%%%%%%%%%%%%%%%%%%%%%
\section{Fit to deep inelastic scattering data}
\label{sec:fit}

We consider the SIDIS process 
$e(l)+N^\uparrow(P) \rightarrow e(l')+ \pi^+(P_1)+ \pi^-(P_2)+ X$,  where $P$ 
is the momentum of the nucleon target with mass $M$, $l,l'$ are the lepton 
momenta before and after the scattering and $q=l-l'$ is the space-like momentum 
transferred to the target. The final pions, with mass $m_\pi = 0.14$ GeV and 
momenta $P_1$ and $P_2$, have invariant mass $M_h$ (which we consider as much 
smaller than the hard scale $Q^2=-q^2 \geq 0$ of the SIDIS process). We 
introduce the pair total momentum $P_h = P_1 + P_2$ and relative momentum 
$R = (P_1-P_2)/2$. Using the traditional Sudakov representation of a 4-momentum 
$a$ in terms of its light-cone components $a^\pm = (a^0 \pm a^3)/\sqrt{2}$ and 
transverse spatial components $\bm{a}_{\sT}$, we define the light-cone 
fractions $x= p^+/P^+$ and $z=P_h^-/k^-$, where $p$ and $k=p+q$ are the momenta 
of the parton before and after the hard vertex, respectively. 

In this process, the following asymmetry can be measured (for the precise
definition we refer to Refs.~\cite{Bacchetta:2006un,Airapetian:2008sk}), 
\begin{equation} 
A_{UT}^{\sin(\phi_R^{} + \phi_S^{})\,\sin\theta}(x,y,z,M_h^2) =
- \frac{\frac{1-y-y^2\,\g^2/4}{x\,y^2\,(1+\g^2)}\,
        \left( 1+\frac{\g^2}{2x}\right)}
       {\frac{1-y+y^2/2 + y^2\,\g^2/4}{x\,y^2\,(1+\g^2)}\,
        \left( 1+\frac{\g^2}{2x}\right)}\,
\frac{|\bm{R}|}{M_h}\,  
\frac{\sum_q e_q^2\,h_1^q(x)\ H_{1,q}^{\open sp}(z,M_h^2)}
     {\sum_q e_q^2\,f_1^q(x)\  D_{1,q}(z,M_h^2)} \; , 
\label{eq:asydis}
\end{equation} 
where $y=P\cdot q/P\cdot l$ is related to the fraction of beam energy 
transferred to the hadronic system, $\g =2Mx/Q$, $f_1^q$ and $h_1^q$ are the 
unpolarized and transversely polarized parton distributions, respectively, and
\begin{equation}
|\bm{R}| = \frac{M_h}{2}\,\sqrt{1-\frac{4m_\pi^2}{M_h^2}} \; .
\label{eq:|R|}
\end{equation}

The spin asymmetry~(\ref{eq:asydis}) is related to an asymmetric modulation of 
pion pairs in the angles $\phi_S^{}$ and $\phi_R^{}$, which represent the 
azimuthal orientation with respect to the scattering plane of the target 
transverse polarization and of the plane containing the pion pair momenta, 
respectively (see Ref.~\cite{Bacchetta:2006un} for a precise definition, which 
is consistent with the Trento conventions~\cite{Bacchetta:2004jz}). 

The polar angle $\theta$ describes the orientation of $P_1$, in the 
center-of-mass frame of the two pions, with respect to the direction of $P_h$ 
in the lab frame. DiFF depend upon $\cos\theta$ via the light-cone fraction 
$\zeta = 2R_{}^-/P_h^- = 2 \cos\theta |\bm{R}|/M_h$, which describes how the 
total momentum of the pair is split between the two 
pions~\cite{Bacchetta:2006un}. DiFF can be expanded in terms of Legendre 
polynomials of $\cos\theta$ and the expansion can be reasonably truncated to 
include only the $s$ and $p$ relative partial waves of the pion pair, since 
their invariant mass is small (typically $M_h \lesssim 1$ GeV). At leading 
twist and leading order in $\alpha_s$, the spin asymmetry of 
Eq.~(\ref{eq:asydis}) contains only $D_{1,q}^{} = D_{1,q}^s + D_{1,q}^p$, which 
includes the diagonal pure $s$- and $p$-wave contributions, and 
$H_{1,q}^{\open sp}$, which originates from the interference between 
them~\cite{Bacchetta:2002ux,Bacchetta:2006un}. Subleading-twist terms have 
different azimuthal dependences~\cite{Bacchetta:2003vn}. 

The polarized DiFF $H_{1,q}^{\open sp}$ represents the chiral-odd partner to 
isolate the transversity distribution $h_1^q$ in the spin asymmetry of 
Eq.~(\ref{eq:asydis}). It describes the interference between the fragmentations 
of transversely polarized quarks into pairs of pions in relative $s$ and $p$ 
waves~\cite{Bacchetta:2002ux}. Together with $D_{1,q}^{}$, it was analytically 
calculated in a spectator-model framework for the first time in 
Ref.~\cite{Radici:2001na}, and later in a refined 
version~\cite{Bacchetta:2006un}. The analytical expressions for 
$D_{1,q}^{} \equiv D_{1,oo}^{}$ and 
$H_{1,q}^{\open sp} \equiv H_{1,ot}^{\open}$ can be found in Eqs.~(23) and (26) 
of Ref.~\cite{Bacchetta:2006un}, respectively. The model parameters were fixed 
by adjusting $D_{1,q}^{}$ to the output of the PYTHIA event generator tuned to 
the SIDIS kinematics at HERMES~\cite{Airapetian:2008sk}; their values are 
listed in Eqs.~(32-35) of Ref.~\cite{Bacchetta:2006un}. Note that the spectator 
model by construction gives 
$D_{1,u}^{} = D_{1,d}^{} = D_{1,\bar{u}}^{} = D_{1,\bar{d}}^{}$ and 
$H_{1,u}^{\open sp} = - H_{1,d}^{\open sp} = - H_{1,\bar{u}}^{\open sp} = 
H_{1,\bar{d}}^{\open sp}$.

The calculated spin asymmetry follows the same trend of the data. In particular, 
the shape of the invariant mass dependence is dominated by a resonance peak at 
$M_h \approx m_\rho$, which is due to the interference between a background 
production of pion pairs in $s$ wave and the $p-$wave component dominated by 
the decay $\rho \to \pi \pi$ of the $\rho$ resonance. Similarly, the model 
displays also another broad peak at $M_h\approx 0.5$ GeV due to the 
$\omega \to \pi \pi \pi$ resonant channel. Both predictions and data show no 
sign change in $A_{UT}^{\sin(\phi_R^{} + \phi_S^{})\,\sin\theta}$ as a function 
of $M_h$ around $M_h \approx m_\rho$, contrary to what was predicted in 
Ref.~\cite{Jaffe:1998hf}. However, the results of Ref.~\cite{Bacchetta:2006un} 
systematically overpredict the experimental data at least by a factor of 
two~\cite{Airapetian:2008sk}. 

To correctly reproduce the size of the asymmetry, we multiply the model 
prediction of $H_{1,q}^{\open sp}$ by an extra parameter $\alpha$, while we use 
the model prediction for $D_{1,q}^{}$ without further changes, since its 
parameters have been already fitted to reproduce the unpolarized cross section, 
as predicted by PHYTIA. We use also the GRV98 LO parametrization for $f_1^q$ at 
the HERMES scale $Q^2=2.5$ GeV$^2$. For $h_1^q$, we use the recently extracted 
parametrization from Ref.~\cite{Anselmino:2008sj}, whose central value is 
basically the same as the former one from Ref.~\cite{Anselmino:2007fs} in the 
region $x\lesssim 0.2$ of interest here. The asymmetry is calculated by 
averaging the numerator and denominator of 
$A_{UT}^{\sin(\phi_R^{} + \phi_S^{})\,\sin\theta}$ in each experimental bin, 
while integrating in turns all remaining variables in the ranges
\begin{align}
0.023&<x<0.4,
&
0&<z<0.99,
&
0.5\; {\rm GeV}&<M_h< 1\; {\rm GeV}.
\end{align}
The variable $y$ is always integrated in the range
\begin{equation}
y_{\text{min}} = \text{Max}\bigl[0.1,\, 
Q^2_{\text{min}}/\bigl( x(s-M^2) \bigr),\, 
(W^2_{\text{min}}-M^2)/\bigl((1-x)(s-M^2)\bigr)  \bigr] \; ,
\end{equation}
with $s= 56.2$ GeV$^2$ and $W^2_{\text{min}} = 4$ GeV$^2$.

%%%%%%%%%%%%%%%%%%%%%%%%%%%%%%%%% Fig. 1 %%%%%%%%%%%%%%%%%%%%%%%%%%%%%%%
\begin{figure}[h]
\begin{center}
\includegraphics[height=5cm]{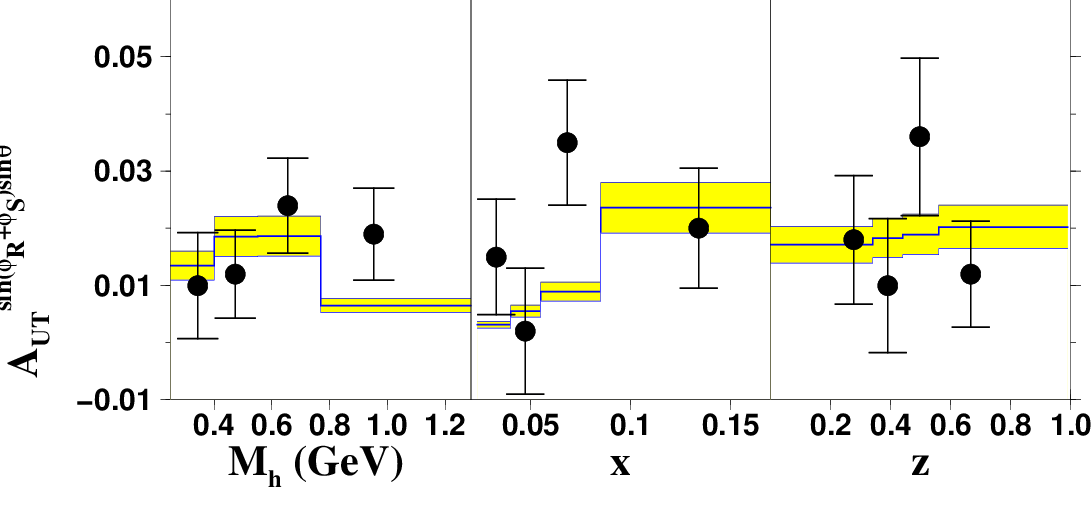} 
\end{center}
\caption{The spin asymmetry for the semi-inclusive production of a pion pair in 
deep-inelastic scattering on a transversely polarized proton, as a function of 
the invariant mass $M_h$ of the pion pair, of the light-cone momentum fraction 
$x$ of the initial parton, of the energy fraction $z$ carried by the pion pair 
with respect to the fragmenting parton. Data from 
Ref.~\protect{\cite{Airapetian:2008sk}}. The uncertainty band is a fit to the 
data based on the DiFF spectator model of 
Ref.~\protect{\cite{Bacchetta:2006un}} and on the $h_1$ parametrization of 
Ref.~\protect{\cite{Anselmino:2008sj}}. }
\label{fig:fit}
\end{figure}

%%%%%%%%%%%%%%%%%%%%%%%%%%%%%%%%%%%%%%%%%%%%%%%%%%%%%%%%%%%%%%%%%%%%%%%%%

The best value for the parameter $\alpha$ is found by means of a $\chi^2$ fit; 
the results are shown in Fig.~\ref{fig:fit}. We took into consideration the 
experimental errors by adding in quadrature the statistical and systematic 
errors (the error bands in Fig.~\ref{fig:fit} represent such a sum). We did not 
include the small theoretical errors coming from the uncertainty on the other 
model parameters of the fragmentation functions, from the uncertainties of the 
parton distribution functions, and from the choice of factorization scale. 
% produced by neglecting higher-order corrections in the evolution. 
The best-fit value of our reduction parameter turns out to be 
$\alpha = 0.32 \pm 0.06$ corresponding to $\chi^2/{\rm d.o.f.} = 1.24$. In 
summary, the model calculation of the $H_{1,q}^{\open sp}$ function has to be 
reduced by a factor 3 to reproduce the HERMES data, if the transversity from 
Ref.~\cite{Anselmino:2008sj} is used.

%%%%%%%%%%%%%%%%%%%%%%%%%%%%%%%%%%%%%%%%%%%%%%%%%%%%%%%%%%%%

\section{Evolution of Dihadron Fragmentation Functions}
\label{sec:evolution}

In order to predict the azimuthal asymmetry in the distribution of two pion 
pairs produced in $e^+e^-$ annihilation, we need to evolve the DiFF 
$D_{1,q}^{}$ and $H_{1,q}^{\open sp}$ from the HERMES scale to the BELLE scale.

DiFF usually depend on $z$, $\zeta = 2 \cos\theta |\bm{R}|/M_h$ [or, 
alternatively, on $z_1 = z (1+\zeta )/2$, $z_2 = z (1-\zeta )/2$], and on 
$R_{\sT}^2$, which is connected to the pair invariant mass 
by~\cite{Ceccopieri:2007ip}
\begin{equation}
R_{\sT}^2 = \frac{(P_{1T} - P_{2T})^2}{4} = \frac{z_1 z_2}{z_1+z_2}\,\left[ 
\frac{M_h^2}{z_1+z_2}-\frac{M_1^2}{z_1}-\frac{M_2^2}{z_2}\right] \; .
\label{eq:RT}
\end{equation}
The further dependence on the scale $Q^2$ of the process is described by usual 
DGLAP evolution equations; at LL, they read~\cite{Ceccopieri:2007ip}
\begin{equation}
\frac{d}{d\mathrm{log} Q^2}\, D_q^{}(z_1,z_2,R_{\sT}^2,Q^2) =  
\frac{\alpha_s(Q^2)}{2\pi}\,\int_{z_1+z_2}^1 \frac{du}{u^2}\,
D_{q'}^{}\left( \frac{z_1}{u},\frac{z_2}{u},R_{\sT}^2,Q^2 \right)\, P_{q'q}(u) 
\; , 
\label{eq:dglap}
\end{equation}
where $P(u)$ are the usual leading-order splitting 
functions~\cite{Altarelli:1977zs}. A similar equation holds for $H_q^{\open}$ 
involving the splitting functions $\delta P(u)$ for transversely polarized 
partons~\cite{Stratmann:2001pt,Artru:1990zv} (see also the Appendix of 
Ref.~\cite{Ceccopieri:2007ip}, for convenience). 

The same strategy can be applied to study evolution of single components of 
extended DiFF in the expansion in relative partial waves of the pion pair. In 
fact, Eq.~(\ref{eq:dglap}) can be rewritten as
\begin{equation}
\frac{d}{d\mathrm{log} Q^2}\, D_{q}^{}(z,\zeta,M_h^2,Q^2) =  
\frac{\alpha_s(Q^2)}{2\pi}\,\int_z^1 \frac{du}{u}\,
D_{q'}^{}\left( \frac{z}{u},\zeta ,M_h^2,Q^2 \right) \, P_{q'q}(u) \; . 
\label{eq:dglap2}
\end{equation}
Note that the evolution kernel affects only the dependence on $z$, leaving 
untouched the dependence on $\zeta$. That is, it affects the dependence on the 
fractional momentum of the pion pair with respect to the hard fragmenting 
parton, but not the dependence on the nonperturbative processes that make the 
fractional momentum split inside the pair itself. The net effect is that 
extended DiFF display evolution equations very similar to the single-hadron 
fragmentation case. Using the above identity 
$\zeta = 2 \cos\theta |\bm{R}|/M_h$, we can again expand both sides of 
Eq.~(\ref{eq:dglap2}) in terms of Legendre functions of $\cos\theta$ and apply 
the evolution kernel to each member of the expansion. By integrating in 
$d\cos\theta$ both sides we come to the final result
\begin{equation}
\frac{d}{d\mathrm{log} Q^2}\, D_{1,q}^{}(z,M_h^2,Q^2) =  
\frac{\alpha_s(Q^2)}{2\pi}\,\int_z^1 \frac{du}{u}\,
D_{1,q'}^{}\left( \frac{z}{u},M_h^2,Q^2 \right) \, P_{q'q}(u) \; , 
\label{eq:dglap3}
\end{equation}
that involves the DGLAP evolution of the single diagonal component 
$D_{1,q}^{}=D_{1,q}^s + D_{1,q}^p$ related to the pure $s$ and $p$ relative 
partial waves of the pion pair. Analogously, we can get an evolution equation 
similar to Eq.~(\ref{eq:dglap3}) for $H_{1,q}^{\open sp}$ provided that $P(u)$ 
is replaced by $\delta P(u)$. 

%%%%%%%%%%%%%%%%%%%%%%%%%%%%%%%%% Fig. 2 %%%%%%%%%%%%%%%%%%%%%%%%%%%%%%%
\begin{figure}[h]
\begin{center}
\includegraphics[height=6cm]{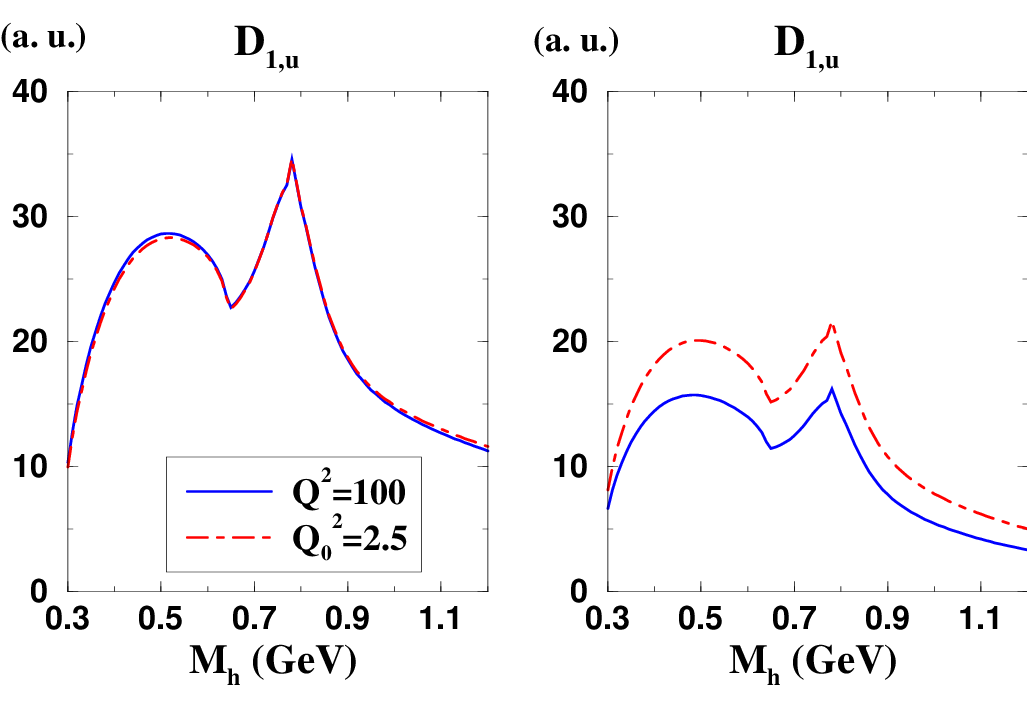} 
\end{center}
\caption{The unpolarized extended DiFF $D_{1,u}^{}(M_h,Q^2)$ in arbitrary 
units, after integrating the $z$ dependence away in the interval $[0.02,1]$ 
(left panel) and $[0.2,1]$ (right panel). Dot-dashed line for $Q^2=2.5$ GeV$^2$ 
at HERMES, solid line for $Q^2=100$ GeV$^2$ at BELLE (see text).}
\label{fig:dglap}
\end{figure}

%%%%%%%%%%%%%%%%%%%%%%%%%%%%%%%%%%%%%%%%%%%%%%%%%%%%%%%%%%%%%%%%%%%%%%%%%

Equation~(\ref{eq:dglap3}) shows that also the dependence on the pair invariant 
mass $M_h$ is not affected by the evolution kernel, as is reasonable, since 
$M_h$ is a scale much lower than $Q^2$. However, in order to get the $M_h$ 
dependence at a different scale $Q^{\prime\,2}\neq Q^2$ it is important to 
completely integrate away the $z$ dependence. Usually, experimental phase 
spaces are limited by the geometry of the apparatus and, in this case, the 
integration in $dz$ is performed in the interval $[z_{\mathrm{min}},1]$ with 
$z_{\mathrm{min}} \neq 0$. In Fig.~\ref{fig:dglap}, we show $D_{1,u}^{}(M_h)$ 
for the up quark at the HERMES scale $Q^2=2.5$ GeV$^2$ (dot-dashed line) and at 
the BELLE scale of $Q^2=100$ GeV$^2$ (solid line). In the left panel, results 
are obtained using $z_{\mathrm{min}}=0.02$, in the right panel with 
$z_{\mathrm{min}}=0.2$ as in the BELLE setup. Since the DGLAP evolution shifts 
the strength at lower $z$ for increasing $Q^2$, cutting the $z$ phase space 
from below makes the final result miss most of the strength and, consequently, 
display a reduced $D_{1,q}^{}(M_h)$. The apparent (and contradictory) effect of 
perturbative evolution on the dependence upon the nonperturbative scale $M_h$ 
in the right panel is actually spurious, and it disappears as soon as the phase 
space for $z$ integration is properly enlarged to include the lowest $z$ for 
$z_{\mathrm{min}}\to 0$, as shown in the left panel. When extracting azimuthal 
(spin) asymmetries, it is, therefore, crucial to keep in mind these features to 
estimate the effect of evolution. 

As a last general comment, we stress that the analysis of evolution is 
facilitated by the fact that azimuthal asymmetries based on the mechanism of 
dihadron fragmentation can be studied using collinear factorization. This 
feature makes them a cleaner observable than the Collins effect, from the 
theoretical point of view. 

In the next section, we compute azimuthal asymmetries for two pion pairs 
production in $e^+ e^-$ annihilations using evolved extended DiFF. The goal is 
to make model predictions at BELLE kinematics, and, at the same time, to 
estimate the evolution effects, both the pure one on the $z$ dependence and the 
spurious one on the $M_h$ dependence, due to the limited experimental phase 
space.

%%%%%%%%%%%%%%%%%%%%%%%%%%%%%%%%% Fig. 3 %%%%%%%%%%%%%%%%%%%%%%%%%%%%%%%
\begin{figure}[h]
\begin{center}
\includegraphics[height=7cm]{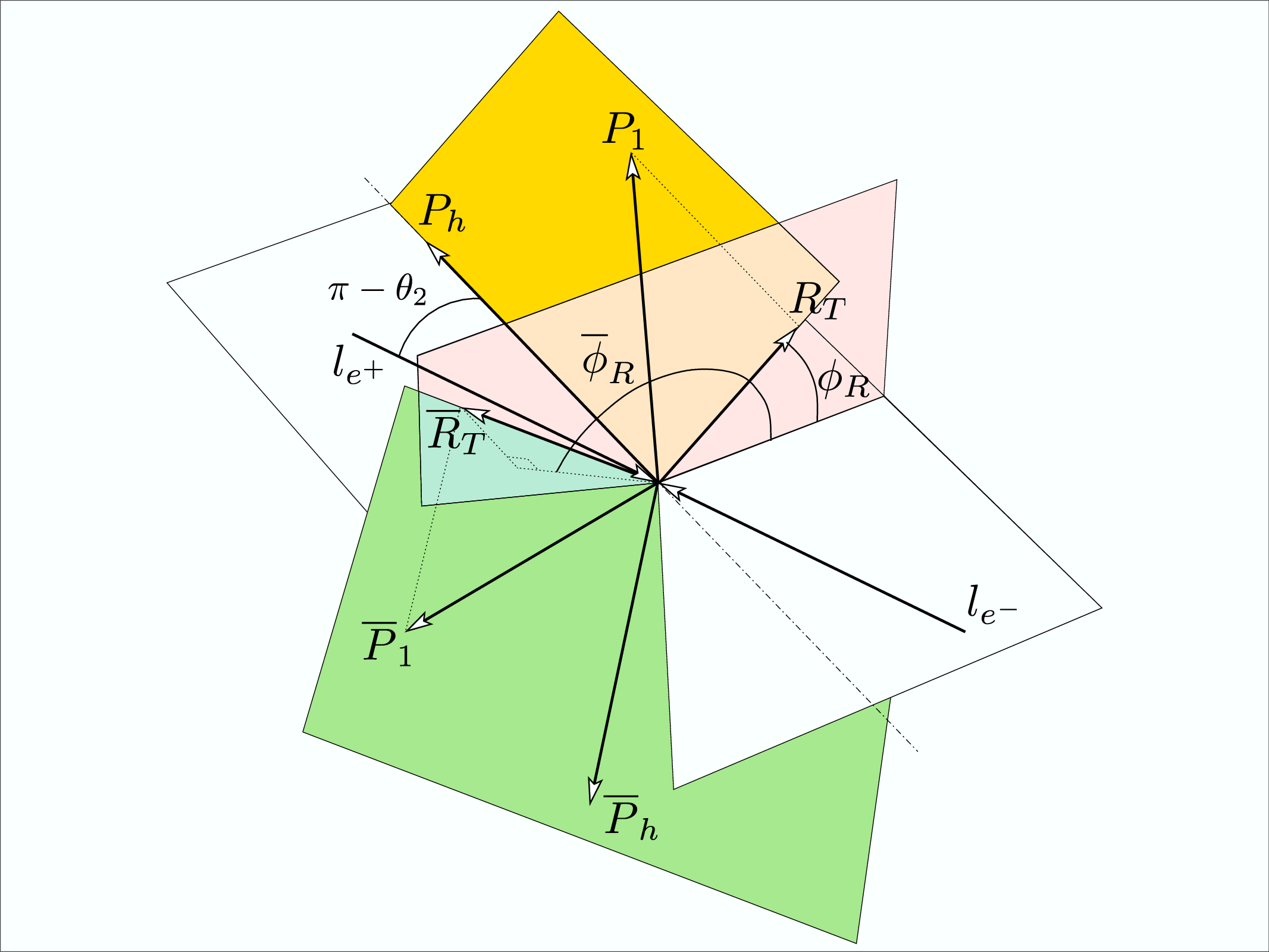} 
\end{center}
\caption{Momenta and angles involved in the process 
$e^+ e^- \rightarrow (\pi^+ \pi^-)_{\rm jet1} (\pi^+ \pi^-)_{\rm jet2} X$.}
\label{fig:kin}
\end{figure}

%%%%%%%%%%%%%%%%%%%%%%%%%%%%%%%%%%%%%%%%%%%%%%%%%%%%%%%%%%%%%%%%%%%%%%%%%

%%%%%%%%%%%%%%%%%%%%%%%%%%%%%%%%%%%%%%%%%%%%%%%%%%%%%%%%%%%%
\section{Predictions for electron-positron annihilation}
\label{sec:predictions}

For the process 
$e^+ e^- \rightarrow (\pi^+ \pi^-)_{\rm jet1} (\pi^+ \pi^-)_{\rm jet2} X$, the 
momentum transfer $q=l+l'$ is time-like, i.e. $q^2=Q^2\geq 0$, with $l,l'$, the 
momenta of the two annihilating fermions. We have now two pairs of pions, one 
originating from a fragmenting parton and the other one from the related 
antiparton. Therefore, we will use in analogy to Sec.~\ref{sec:fit} the 
variables $\phi_R^{}$, $\theta$, $P_1$, $P_2$, $P_h$, $R$, $M_h$, $z$, $\zeta$, 
for one pair, adding the variables $\overline{\phi}_R^{}$, $\overline{\theta}$, 
$\overline{P}_1$, $\overline{P}_2$, $\overline{P}_h$, $\overline{R}$, 
$\overline{M}_h$, $\overline{z}$, $\overline{\zeta}$, for the other
pair. Since we assume that the two pairs belong to two back-to-back jets, we 
have $P_h\cdot \overline{P}_h\approx Q^2$. The momenta and angles involved in 
the description of the process are depicted in Fig.~\ref{fig:kin}. The 
azimuthal angles $\phi_R^{}$ and $\overline{\phi}_R^{}$ are defined by 
\begin{equation}
\begin{split}
\phi_R^{} &= 
\frac{(\bm{l}_{e^+}\times \bm{P}_h)\,\cdot \bm{R}_{\sT}}
     {|(\bm{l}_{e^+}\times \bm{P}_h)\,\cdot \bm{R}_{\sT}|} \, 
\arccos \left( 
        \frac{\bm{l}_{e^+}\times \bm{P}_h}{|\bm{l}_{e^+}\times \bm{P}_h|}
	\cdot 
	\frac{\bm{R}_{\sT}\times \bm{P}_h}{|\bm{R}_{\sT}\times \bm{P}_h|} 
	\right)
\\
\overline{\phi}_R^{} &= 
\frac{(\bm{l}_{e^+}\times \bm{P}_h)\,\cdot \overline{\bm{R}}_{\sT}}
     {|(\bm{l}_{e^+}\times \bm{P}_h)\,\cdot \overline{\bm{R}}_{\sT}|} \, 
\arccos \left( 
        \frac{(\bm{l}_{e^+}\times \bm{P}_h)}{|\bm{l}_{e^+}\times \bm{P}_h|}
        \cdot 
	\frac{(\overline{\bm{R}}_{\sT}\times \bm{P}_h)}
	     {|\overline{\bm{R}}_{\sT}\times \bm{P}_h|} 
	\right)  \; , 
\label{eq:az_angles}
\end{split}
\end{equation}
where $l_{e^+}$ is the momentum of the positron, and 
$\bm{R}_{\sT},\,\overline{\bm{R}}_{\sT}$ indicate the transverse component of 
$\bm{R},\,\overline{\bm{R}}$ with respect to $\bm{P}_h,\,\overline{\bm{P}}_h$, 
respectively. They are measured in the plane identified by 
$\bm{l}_{e^+}\times \bm{P}_h$ and $(\bm{l}_{e^+}\times \bm{P}_h)\times\hat{z}$, 
with $\hat{z} \parallel -\bm{P}_h$ in analogy to the Trento 
conventions~\cite{Bacchetta:2004jz}; this plane is perpendicular to the lepton 
plane identified by $\bm{l}_{e^+}$ and $\hat{P}_h$ (see Fig.~\ref{fig:kin}).
Note that the difference between $\overline{\phi}_R^{}$, as defined in
Eq.~(\ref{eq:az_angles}), and the azimuthal angle of $\overline{\bm{R}}_{\sT}$, 
as measured around $\overline{\bm{P}}_h$, is a higher-twist effect. The 
invariant $y = P_h\cdot l / P_h \cdot q$ is now related, in the lepton 
center-of-mass frame, to the angle 
$\theta_2 = \arccos (\bm{l_{e^+}}\cdot\bm{P}_h / (|\bm{l_{e^+}}|\,|\bm{P}_h|))$ 
by $y = (1+\cos\theta_2)/2$. 

Starting from Eq.~(21) of Ref.~\cite{Boer:2003ya}, we define the so-called 
Artru--Collins azimuthal asymmetry 
$A(\cos\theta_2,z,\overline{z},M_h^2,\overline{M}_h^2)$ as the ratio between 
weighted leading-twist cross sections for the 
$e^+ e^- \rightarrow (\pi^+ \pi^-)_{\rm jet1} (\pi^+ \pi^-)_{\rm jet2} X$ 
process, once integrated upon all variables but 
$\cos\theta_2,z,\overline{z},M_h^2,\overline{M}_h^2$. We define the weighted 
cross section as 
\begin{equation}
\langle w \rangle = \int d\zeta d\overline{\zeta} 
\int d\phi_{R}^{} d\overline{\phi}_R^{} \int d\bm{q}_{\sT} \,  w \, 
\frac{d\sigma}{d\cos \theta_2\, dz\,d\overline{z}\, d\zeta \, d\overline{\zeta}
\, dM_h^2\, d\overline{M}_h^2\, d\phi_R^{}\, d\overline{\phi}_R^{} \, 
d\bm{q}_{\sT}} \; .
\label{eq:wcross}
\end{equation}

The weight of the numerator in the asymmetry is 
$\cos (\phi_R^{}+\overline{\phi}_R^{})$, for the denominator it is just 1. 
Elaborating upon the work of Ref.~\cite{Boer:2003ya}~\footnote{With respect to 
Ref.~\cite{Boer:2003ya}, we use the modified definition $\zeta = 2\xi -1$, the 
integration measure of the parton transverse momentum reads $z^2 d\bm{k}_{\sT}$, 
and the scaling factor $1/(M_1+M_2)$ of the chiral-odd projections of the 
parton-parton correlator is replaced by $1/M_h$ (and similarly for the 
antiparton correlator), in agreement with 
Refs.~\cite{Bacchetta:2002ux,Bacchetta:2006un}.}, we recall the change of 
variables $\zeta = 2 \cos\theta |\bm{R}|/M_h$ (and similarly for 
$\overline{\zeta}$) and we perform an expansion in terms of Legendre functions 
of $\cos\theta$ (and $\cos\overline{\theta}$) by keeping only the $s$- and 
$p$-wave components of the relative partial waves of the pion pair. By further 
integrating upon $d\cos\theta$ and $d\cos\overline{\theta}$, we deduce the 
analogue of Eq.~(32) of Ref.~\cite{Boer:2003ya} for the specific contribution 
of $s$ and $p$ partial waves to the Artru-Collins azimuthal asymmetry, namely
\begin{equation} 
\begin{split} 
A(\cos\theta_2,z,M_h^2,\overline{z},\overline{M}_h^2) &\equiv
\frac{\langle \cos(\phi_R^{}+\overline{\phi}_R^{})\rangle}
     {\langle 1 \rangle}
\\
&= \frac{\sin^2 \theta_2}{1+\cos^2 \theta_2} \, \frac{\pi^2}{32}\,
\frac{|\bm{R}|\,|\overline{\bm{R}}|}{M_h\,\overline{M}_h} \,
\frac{\sum_q e_q^2 \, H_{1,q}^{\open sp}(z,M_h^2)\,
          \overline{H}_{1,q}^{\open sp}(\overline{z},\overline{M}_h^2)}
     {\sum_q e_q^2\, D_{1,q}^{}(z,M_h^2) \,
          \overline{D}_{1,q}^{}(\overline{z},\overline{M}_h^2) } \; .
\label{eq:asye+e-}
\end{split} 
\end{equation} 
The extended DiFF $D_{1,q}^{}$ and $H_{1,q}^{\open sp}$ are the same universal 
functions appearing in the SIDIS spin asymmetry of Eq.~(\ref{eq:asydis}). 
Hence, tuning model predictions for them at BELLE kinematics would help in 
reducing the uncertainty in the extraction of the transversity $h_1$ at the 
HERMES scale. In this strategy, a crucial role is played by evolution. At 
variance with the Collins effect, the dihadron fragmentation mechanism is fully 
collinear, since only $\bm{R}_{\sT}$ matters and $\bm{P}_{h\perp}$ can be
integrated. Hence, evolution equations for extended DiFF are easily under 
control, presently at LL level~\cite{Ceccopieri:2007ip}, and are represented by 
Eq.~(\ref{eq:dglap3}) and its analogue for $H_{1,q}^{\open sp}$.

%%%%%%%%%%%%%%%%%%%%%%%%%%%%%%%%% Fig. 4 %%%%%%%%%%%%%%%%%%%%%%%%%%%%%%%
\begin{figure}[h]
\begin{center}
\includegraphics[height=5.7cm]{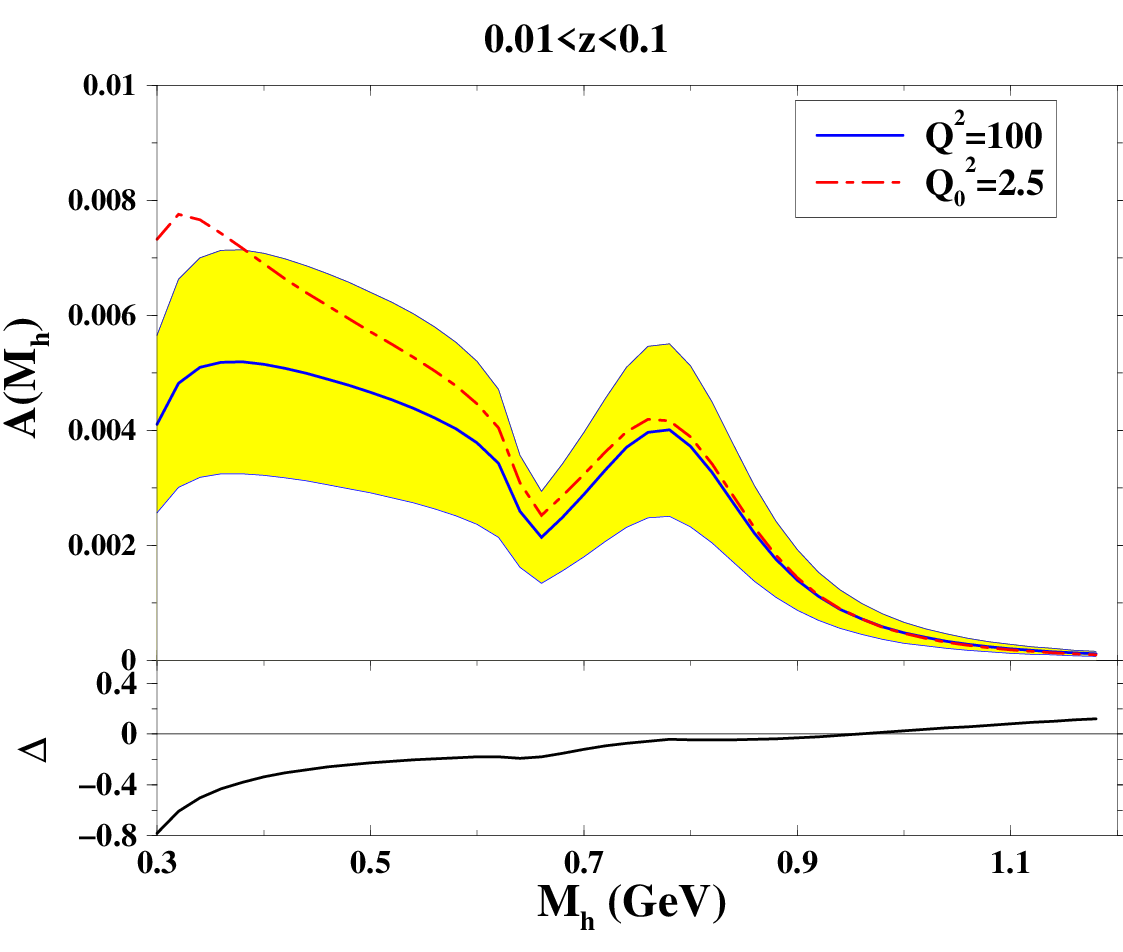}~\hspace{5mm}~\includegraphics[height=5.7cm]{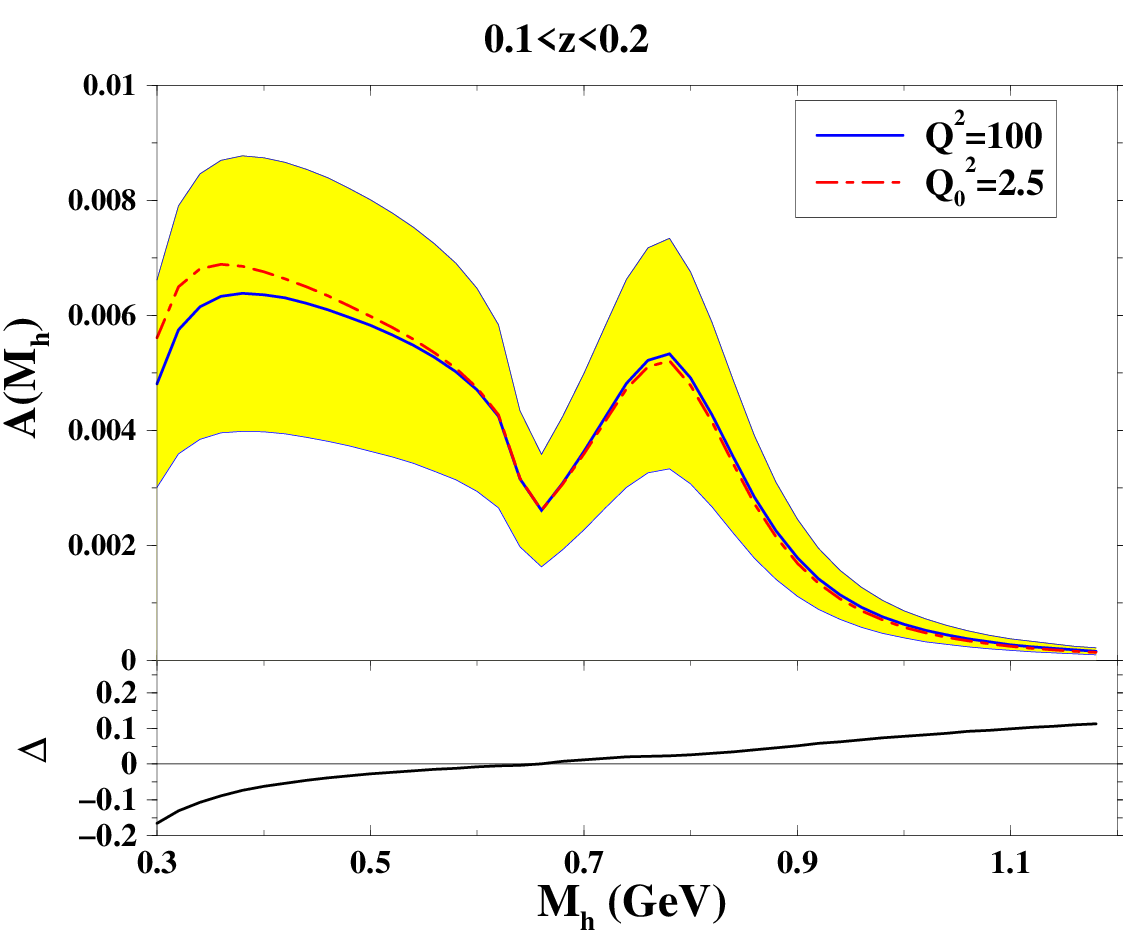} \\[5mm]
\includegraphics[height=5.7cm]{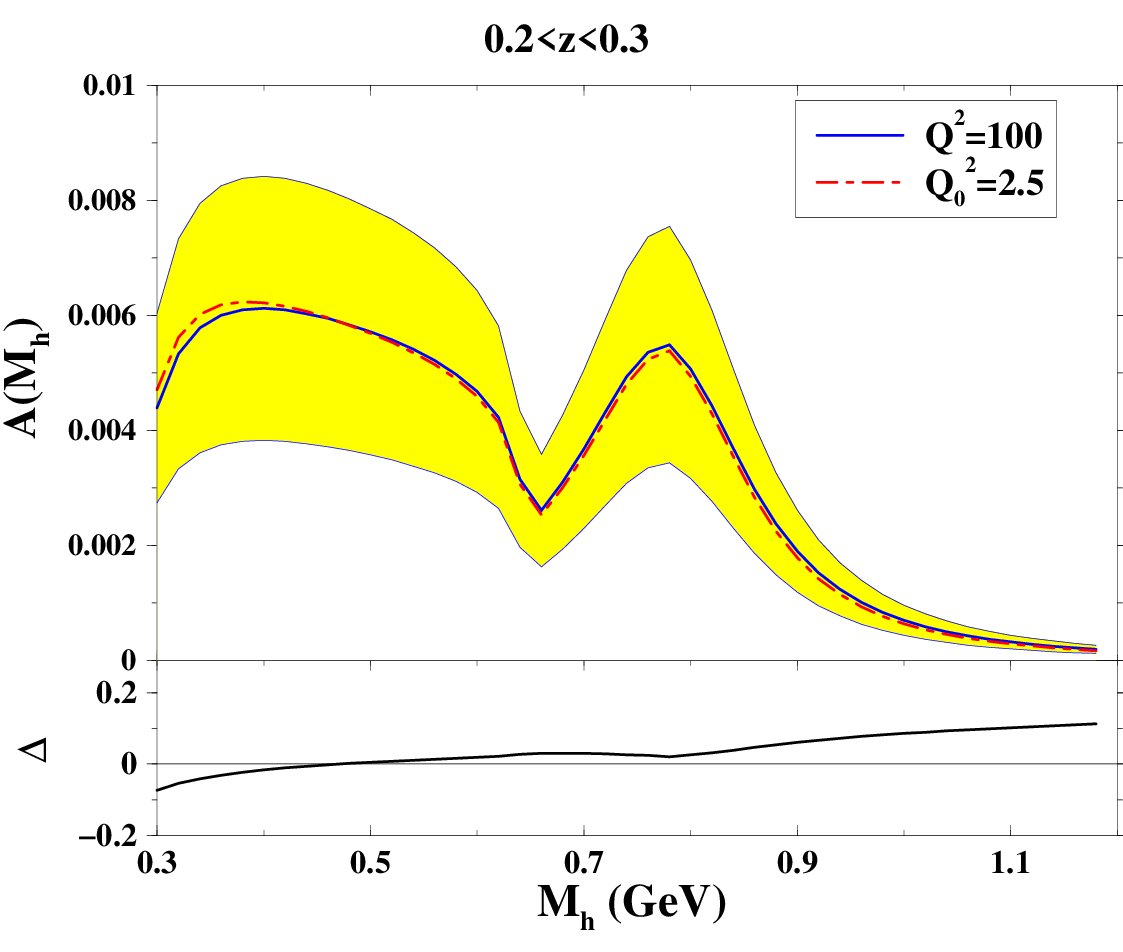}~\hspace{5mm}~\includegraphics[height=5.7cm]{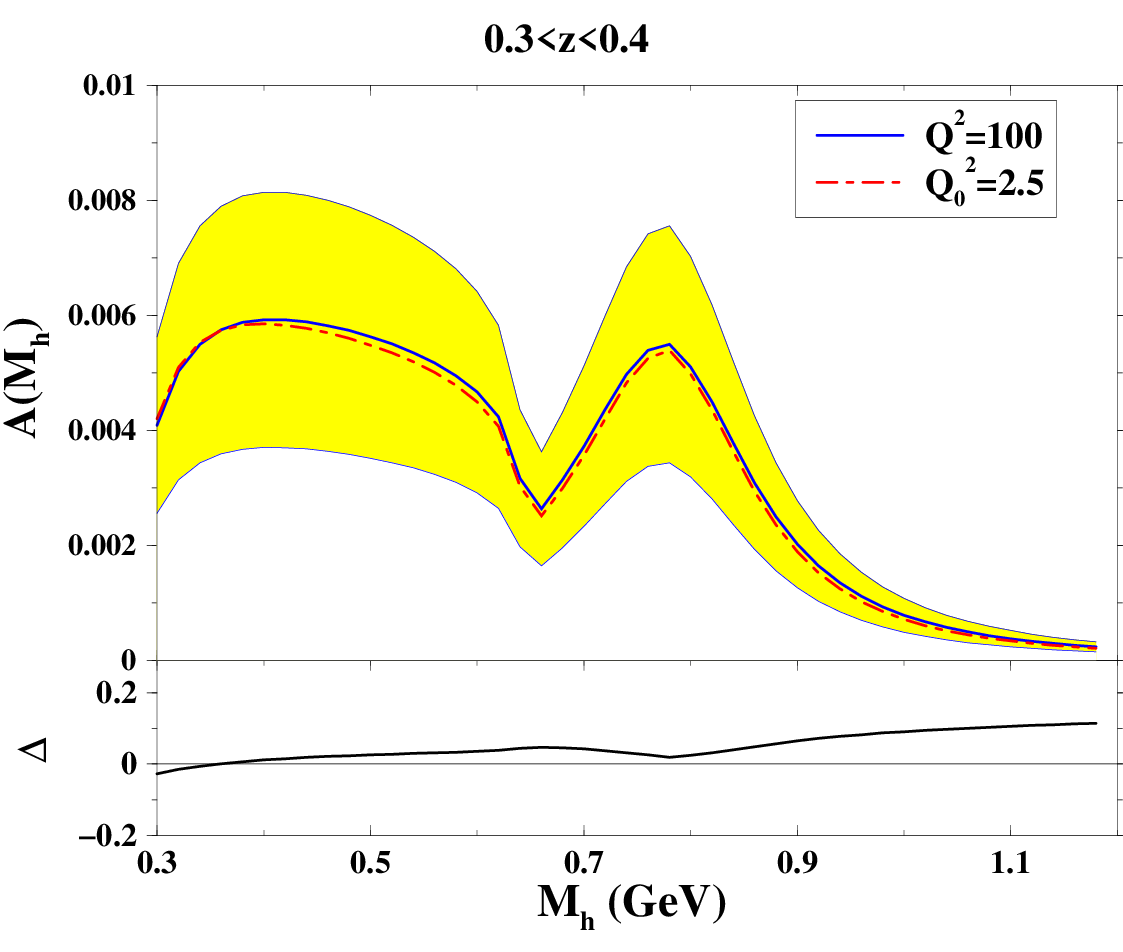} \\[5mm]
\includegraphics[height=5.7cm]{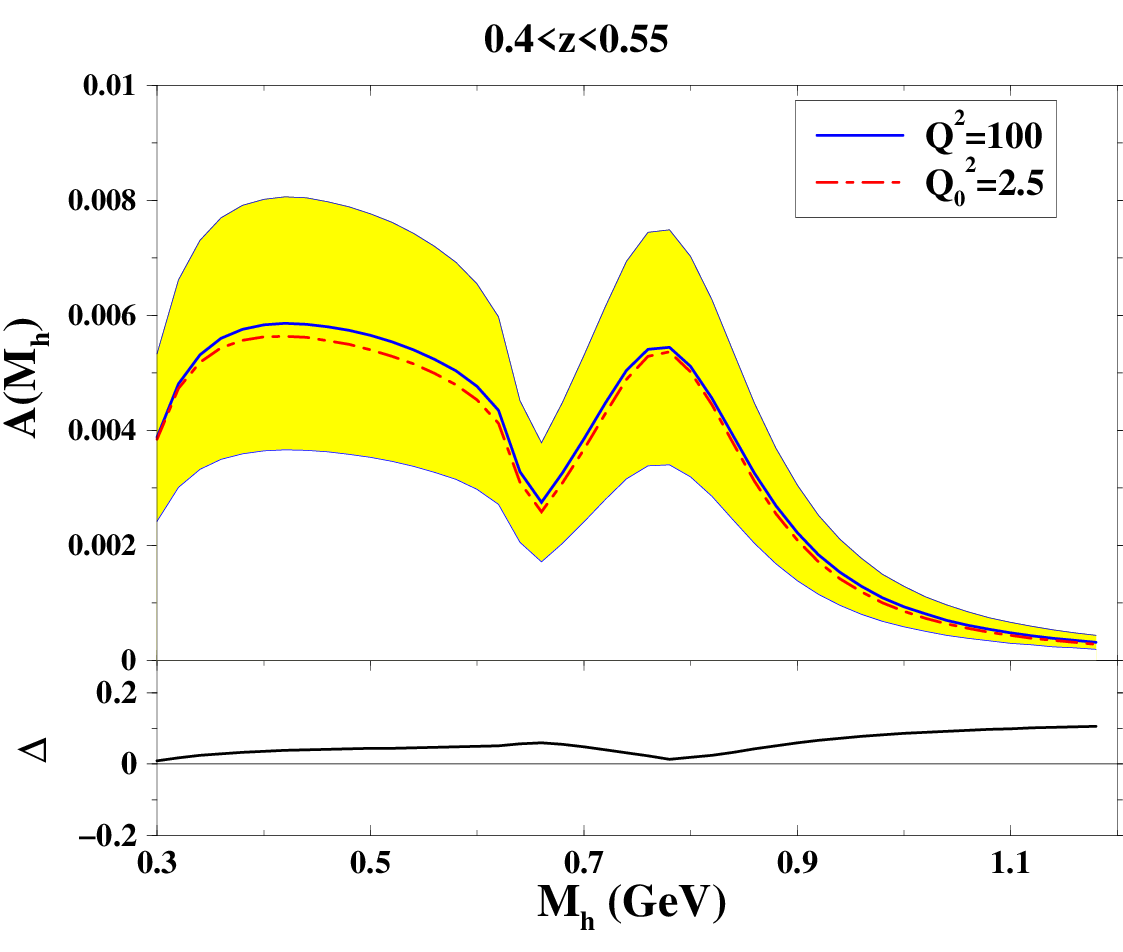}~\hspace{5mm}~\includegraphics[height=5.7cm]{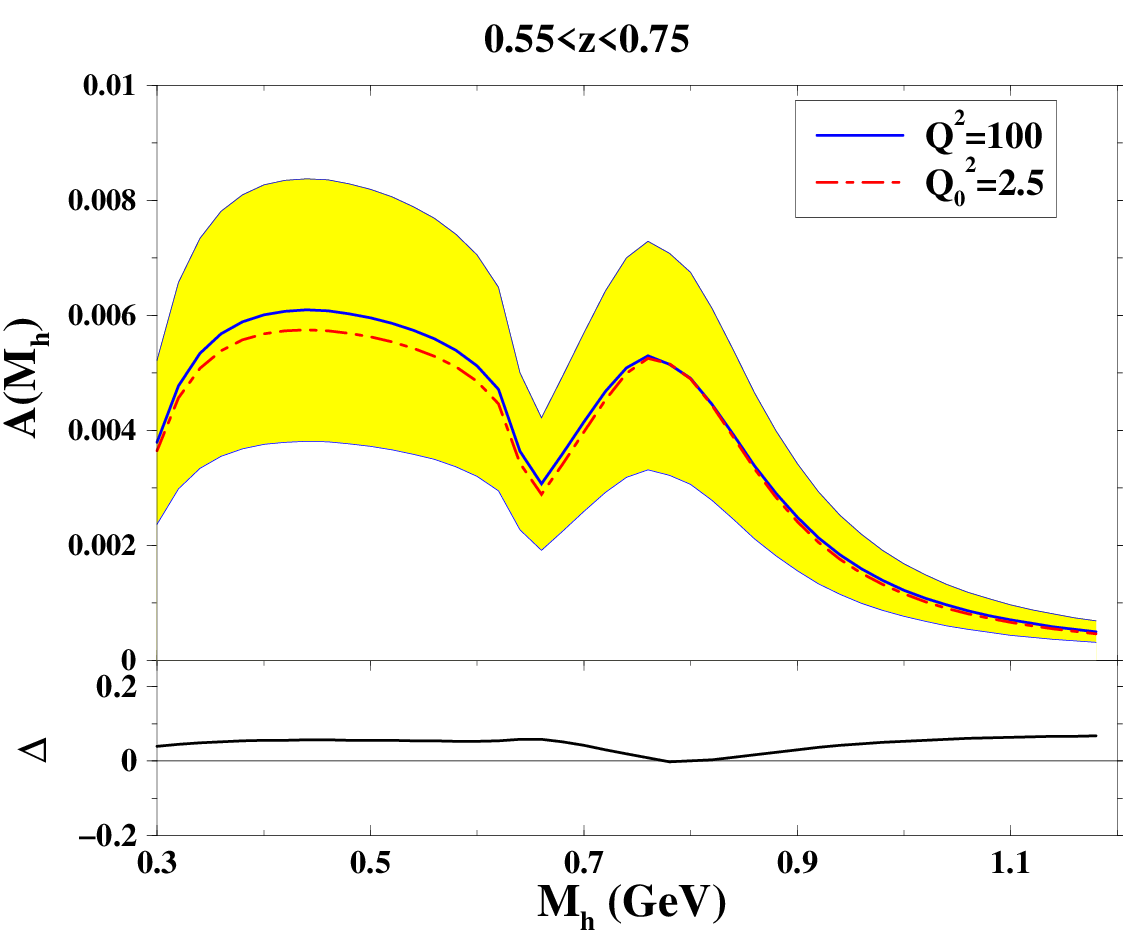}
\end{center}
\caption{The azimuthal asymmetry for two pion pairs production in $e^+ e^-$ 
annihilation as a function of the invariant mass $M_h$ of one pair for the 
indicated bins in its momentum fraction $z$. Notations as in 
Fig.~\protect{\ref{fig:dglap}}. The uncertainty band around the solid line 
originates from the fit error of Fig.~\protect{\ref{fig:fit}} through error 
propagation. For each panel, the lower plot shows the modification factor of 
the final result because of DGLAP evolution (see text).}
\label{fig:AMhz}
\end{figure}

%%%%%%%%%%%%%%%%%%%%%%%%%%%%%%%%%%%%%%%%%%%%%%%%%%%%%%%%%%%%%%%%%%%%%%%%%

In Fig.~\ref{fig:AMhz}, the azimuthal asymmetry of Eq.~(\ref{eq:asye+e-}) is 
displayed as a function of $M_h$ for the $z$ bins $[0.01,0.1]$, $[0.1,0.2]$, 
$[0.2,0.3]$, $[0.3,0.4]$, $[0.4,0.55]$, and $[0.55,0.75]$, after integrating 
upon the other variables $0.4\leq \overline{M}_h\leq 1.2$ GeV, 
$0.2\leq \overline{z}\leq 0.9$, and $-0.6\leq \cos\theta_2\leq 0.9$, according 
to the BELLE experimental phase space. In particular, according to 
Ref.~\cite{Seidl:2008xc} for each bin we have assumed the following coefficient 
\begin{equation}
\frac{\langle \sin^2 \theta_2\rangle}{\langle 1+\cos^2 \theta_2 \rangle} 
\approx 0.7 \; .
\end{equation} 

In each panel, the upper plot shows the $A(z_{\mathrm{bin}},M_h^2)$ at the 
HERMES scale $Q^2=2.5$ GeV$^2$ (dot-dashed line) and at the BELLE scale 
$Q^2=100$ GeV$^2$ (solid line), the latter being supplemented by the 
uncertainty band propagated from the SIDIS fit error of Fig.~\ref{fig:fit}. The 
lower plot shows 
\begin{equation}
\Delta = \frac{A(z_{\mathrm{bin}},M_h^2,Q^2=100) - 
               A(z_{\mathrm{bin}},M_h^2,Q^2=2.5)}
              {A(z_{\mathrm{bin}},M_h^2,Q^2=100)} \; , 
\label{eq:delta}
\end{equation}
i.e. the modification factor of the final result due to the evolution starting 
at the HERMES scale. 

\begin{figure}[h]
\begin{center}
\includegraphics[height=5.7cm]{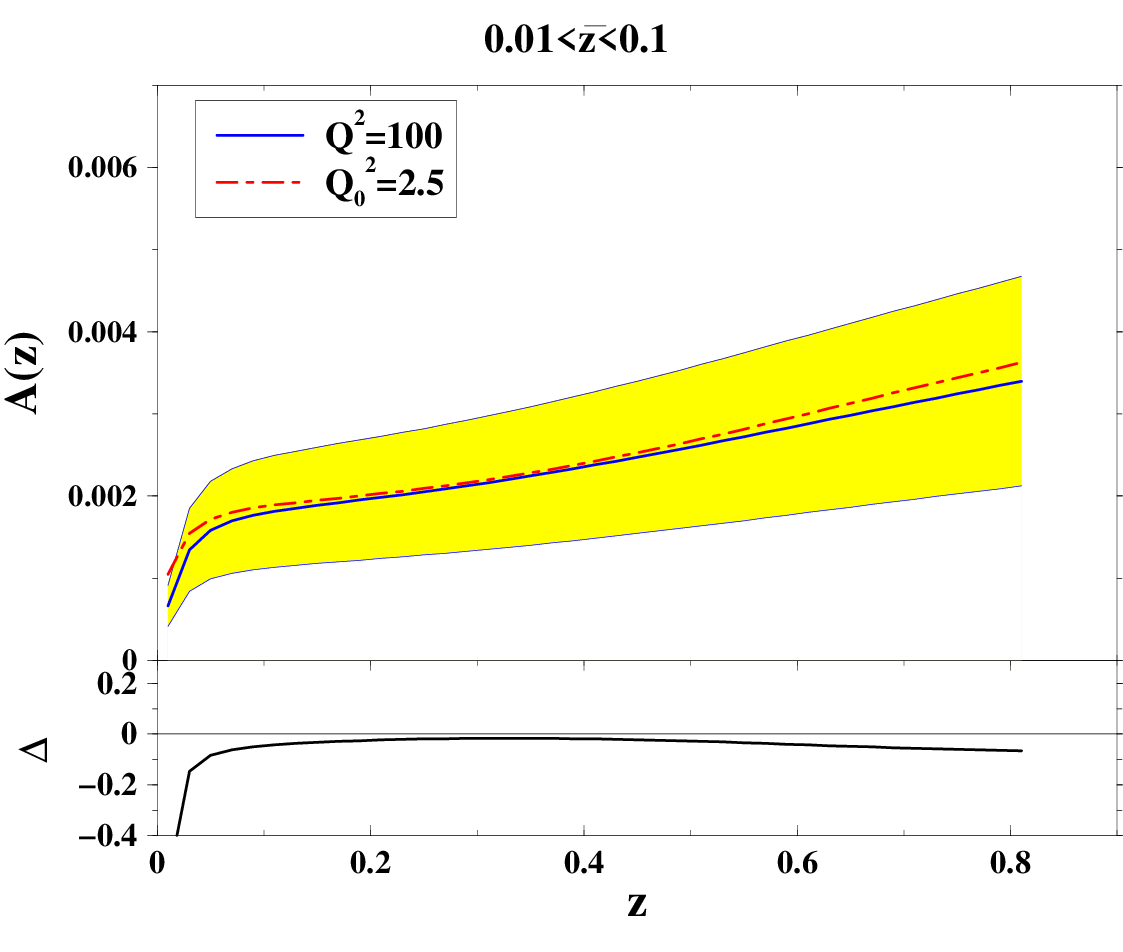}~\hspace{5mm}~\includegraphics[height=5.7cm]{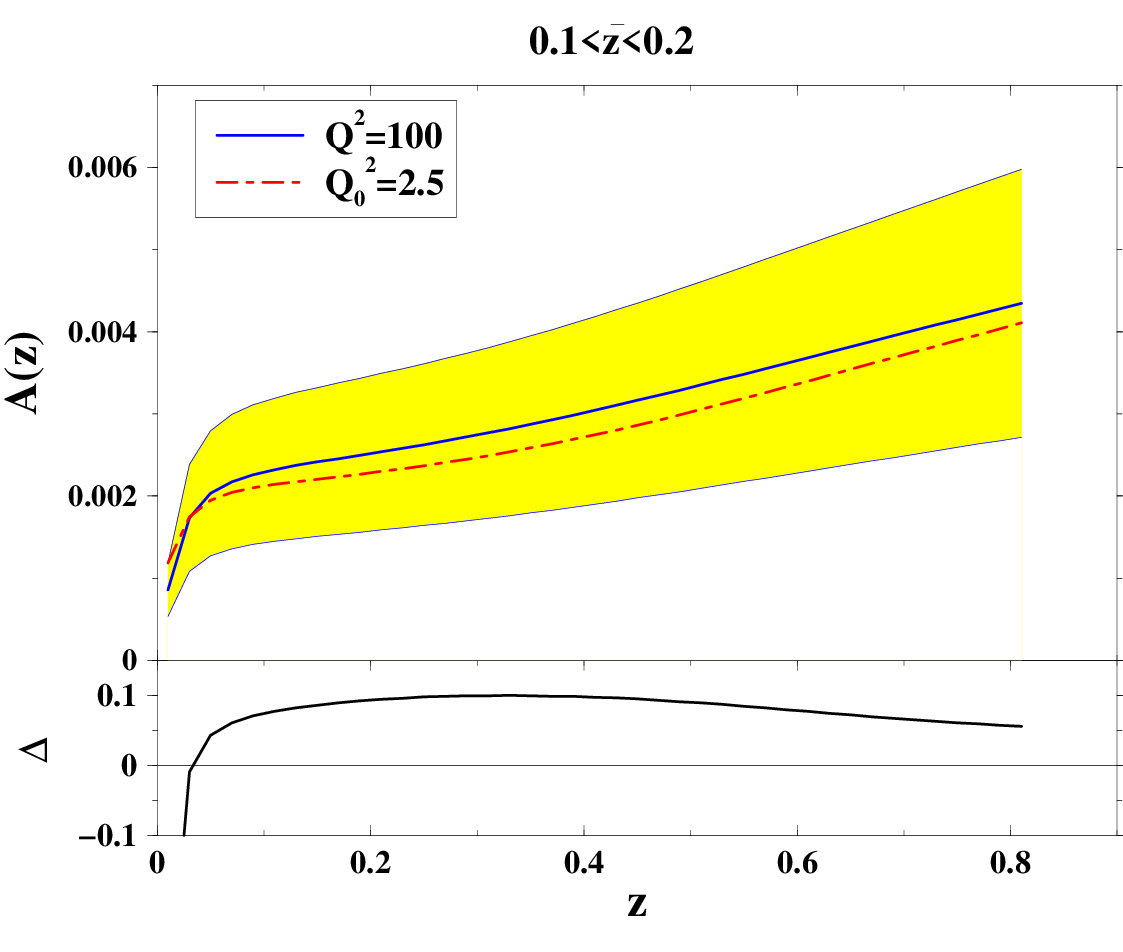} \\[5mm]
\includegraphics[height=5.7cm]{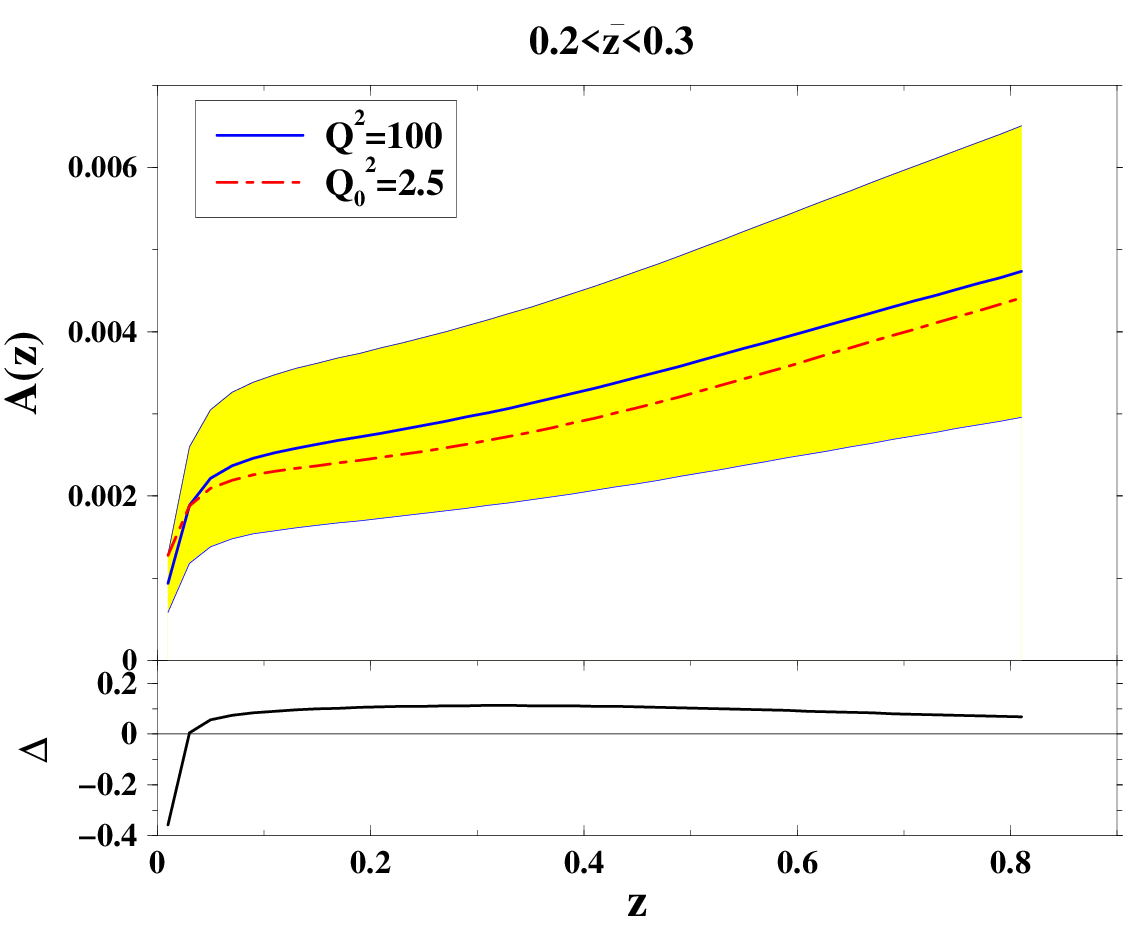}~\hspace{5mm}~\includegraphics[height=5.7cm]{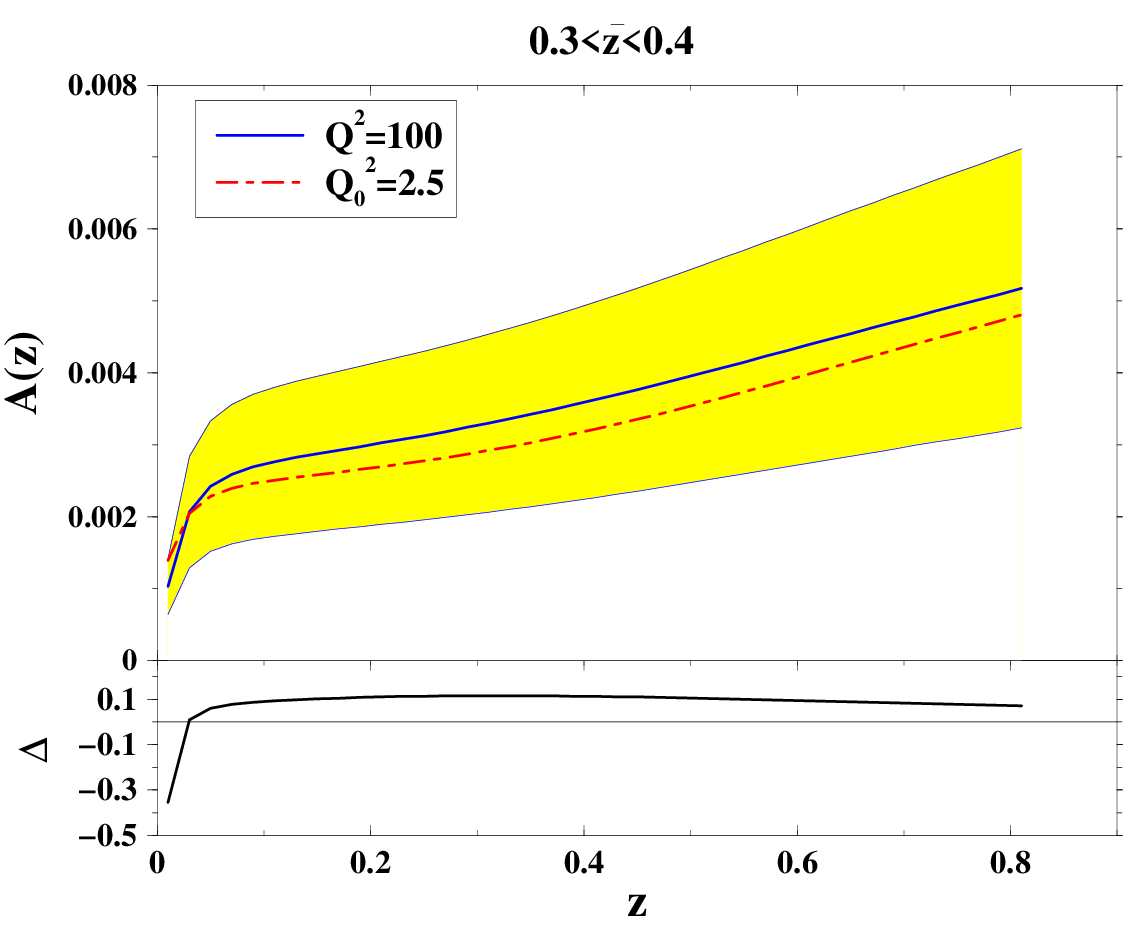} \\[5mm]
\includegraphics[height=5.7cm]{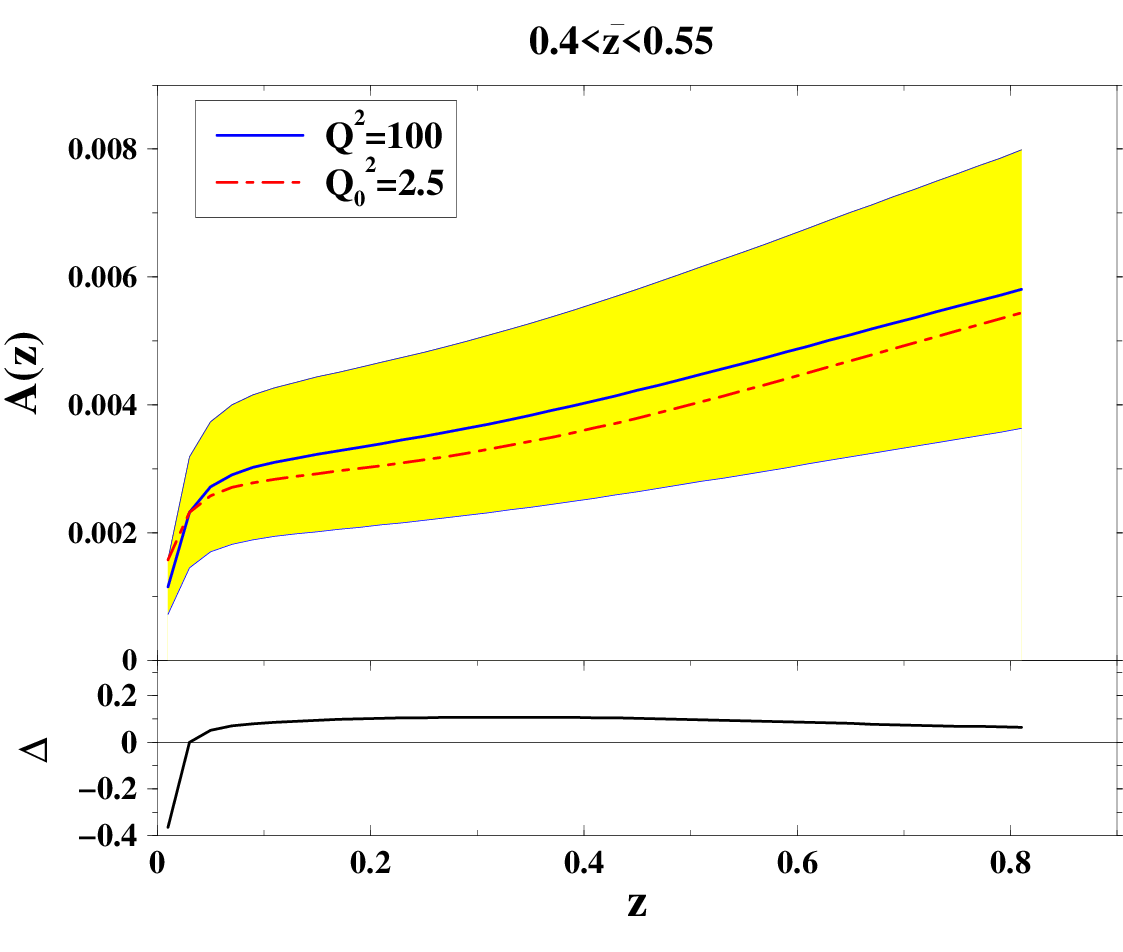}~\hspace{5mm}~\includegraphics[height=5.7cm]{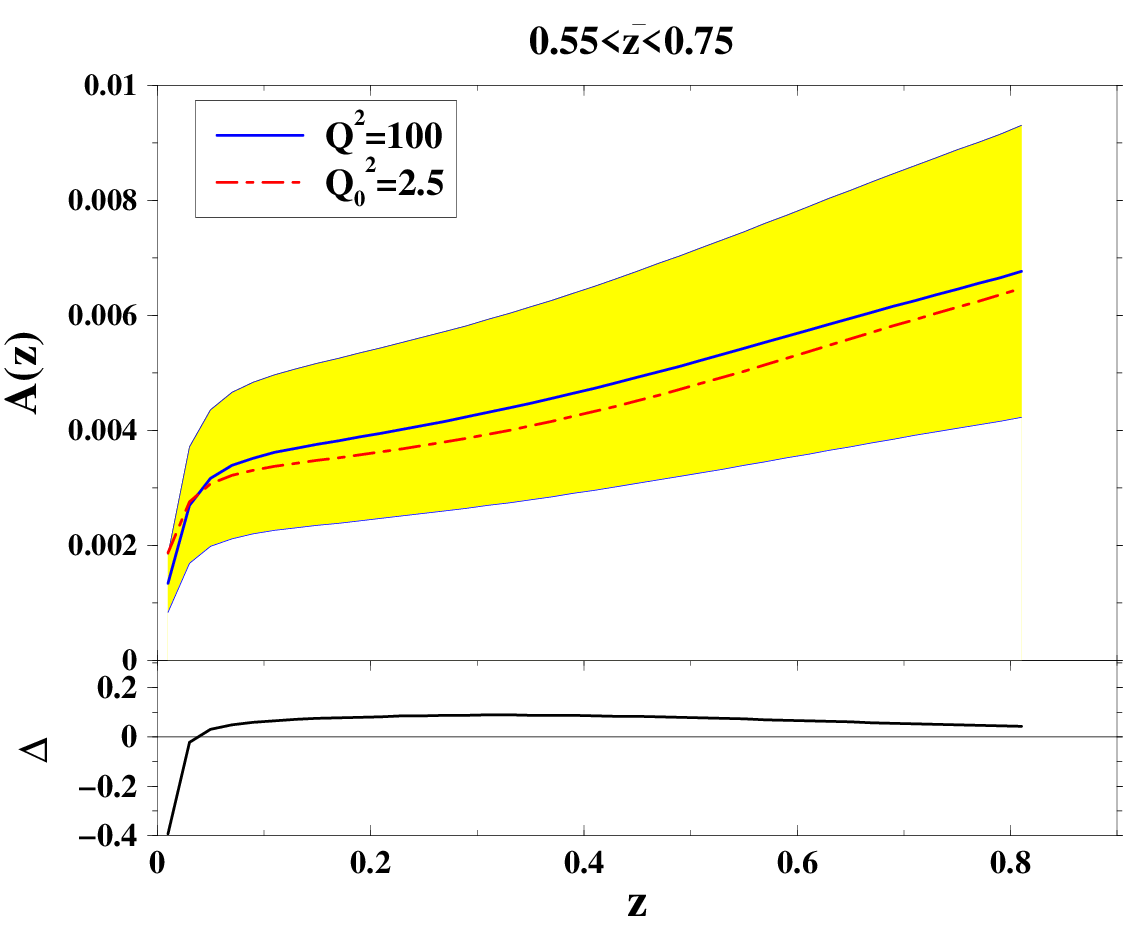}
\end{center}
\caption{Same as in Fig.~\protect{\ref{fig:AMhz}} but as a function of $z$ for 
the indicated $\overline{z}$ bins.}
\label{fig:Azzb}
\end{figure}

Some comments are in order about Fig.~\ref{fig:AMhz}. First of all, the 
absolute size of the Artru--Collins asymmetry is small, reaching at most the 
percent level (see Fig.~\ref{fig:Azzb}). However, it should be within the reach 
of the BELLE experimental capabilities, if compared with the corresponding 
Collins effect for two separated single-hadron 
fragmentations~\cite{Seidl:2008xc}. The error band originates from the 
uncertainty in the size of DiFF due to the fit of the SIDIS data for the spin 
asymmetry of Eq.~(\ref{eq:asydis}). This error band is always much larger than 
the effects due to evolution. 

\begin{figure}[h]
\begin{center}
\includegraphics[height=6cm]{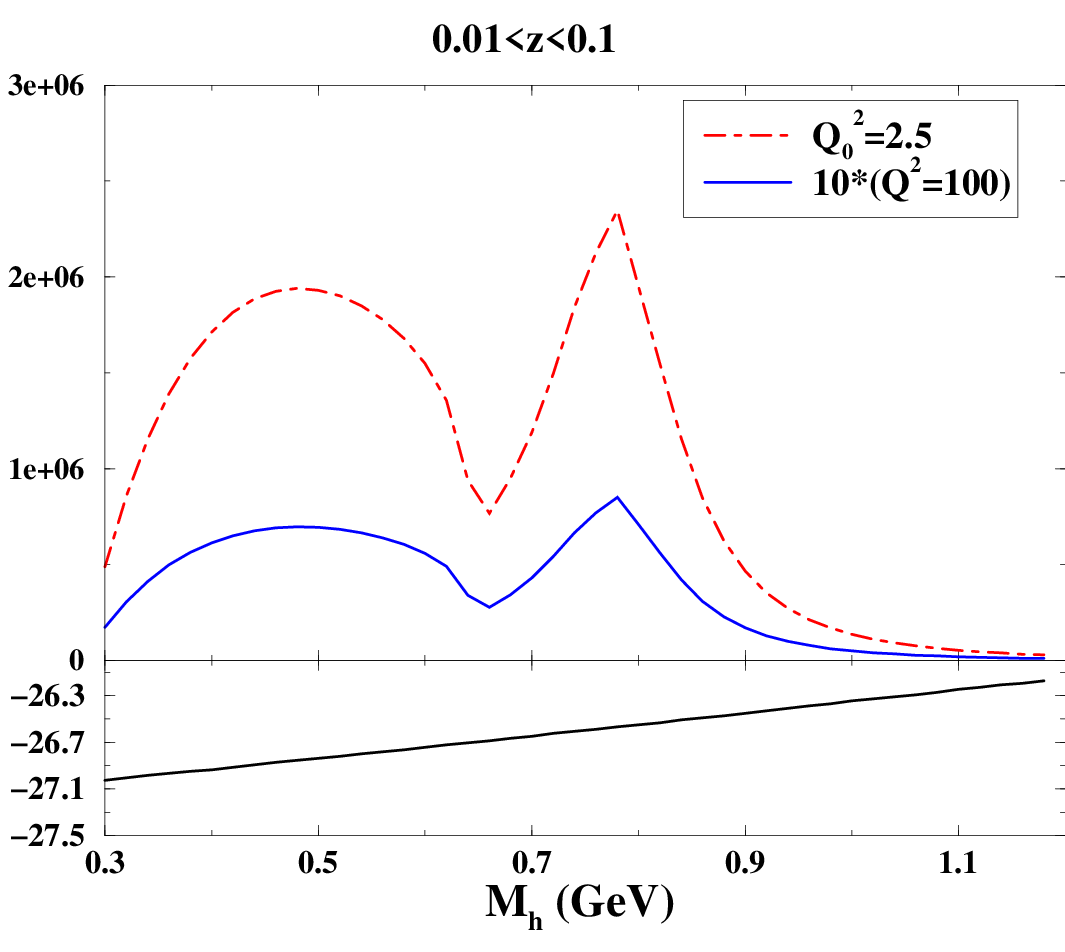}~\hspace{5mm}~\includegraphics[height=6cm]{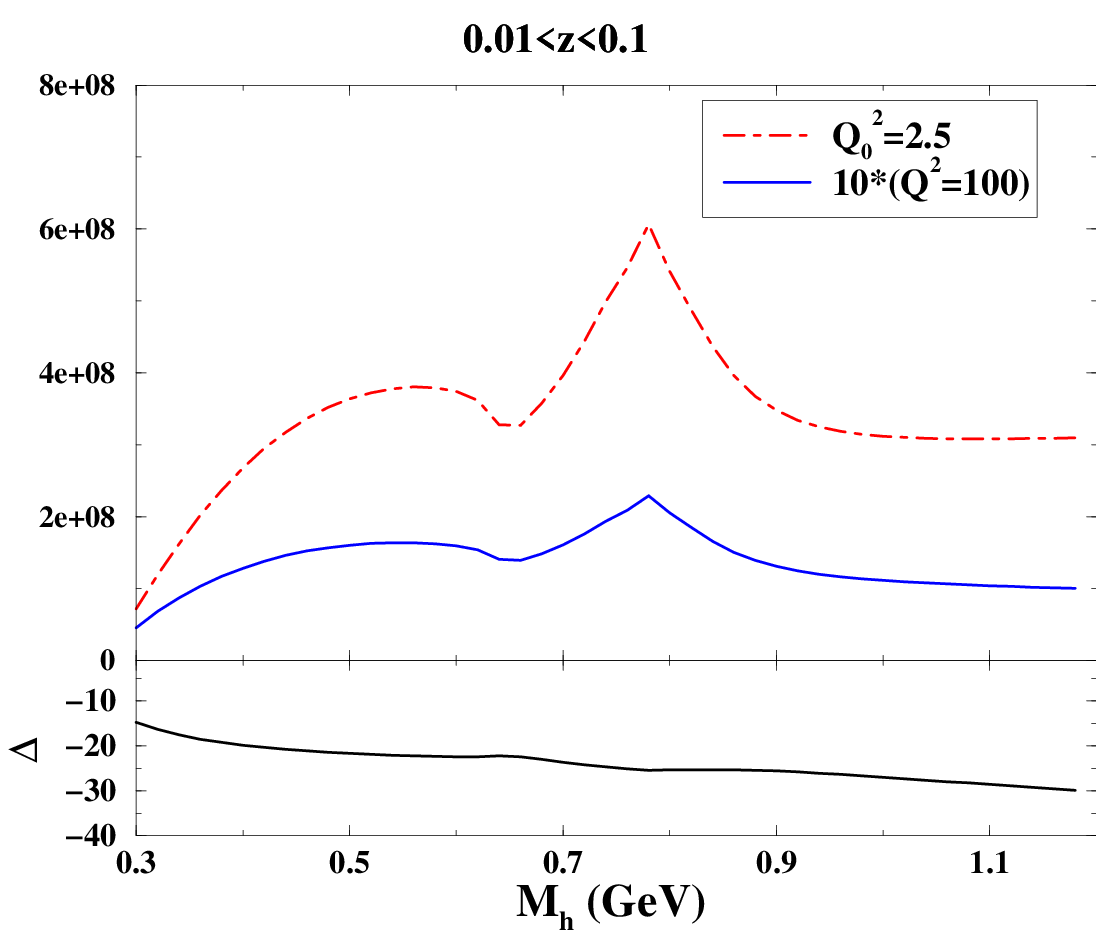} \\[5mm]
\includegraphics[height=6cm]{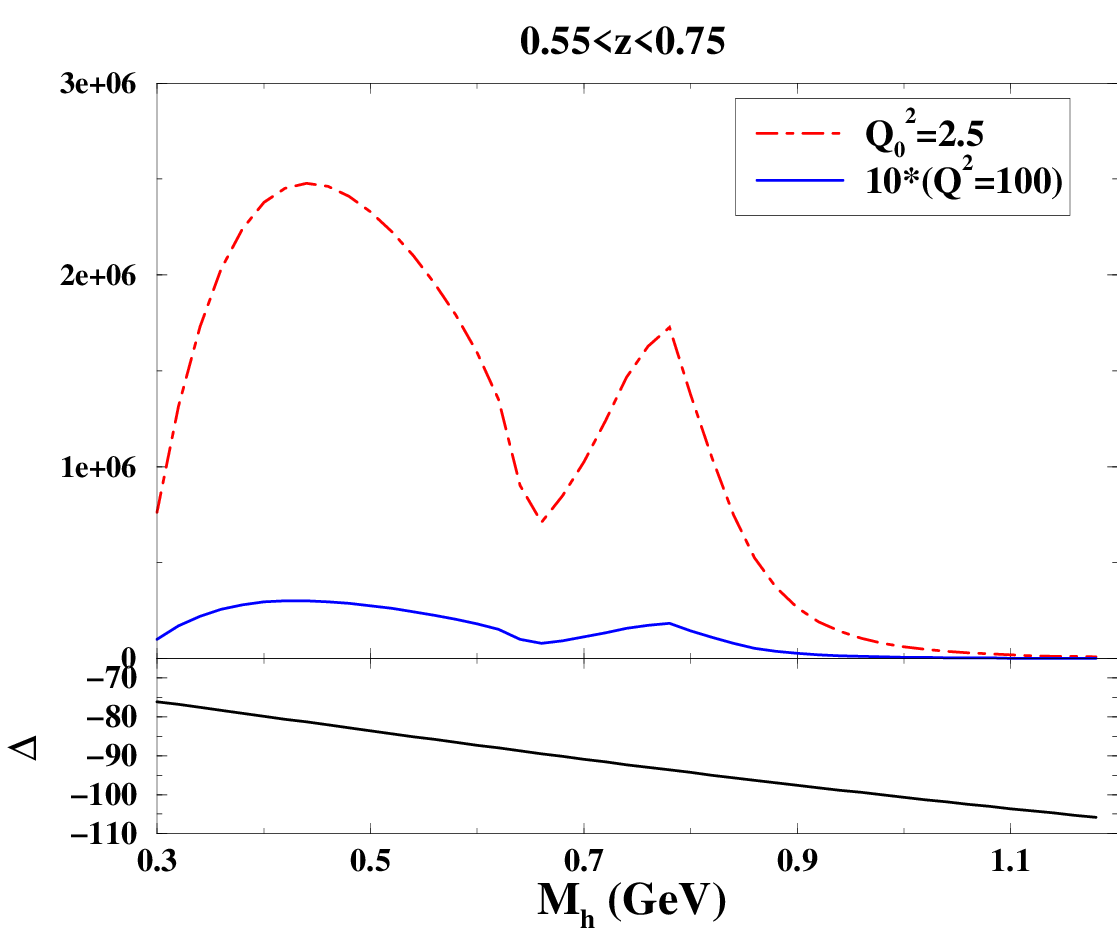}~\hspace{5mm}~\includegraphics[height=6cm]{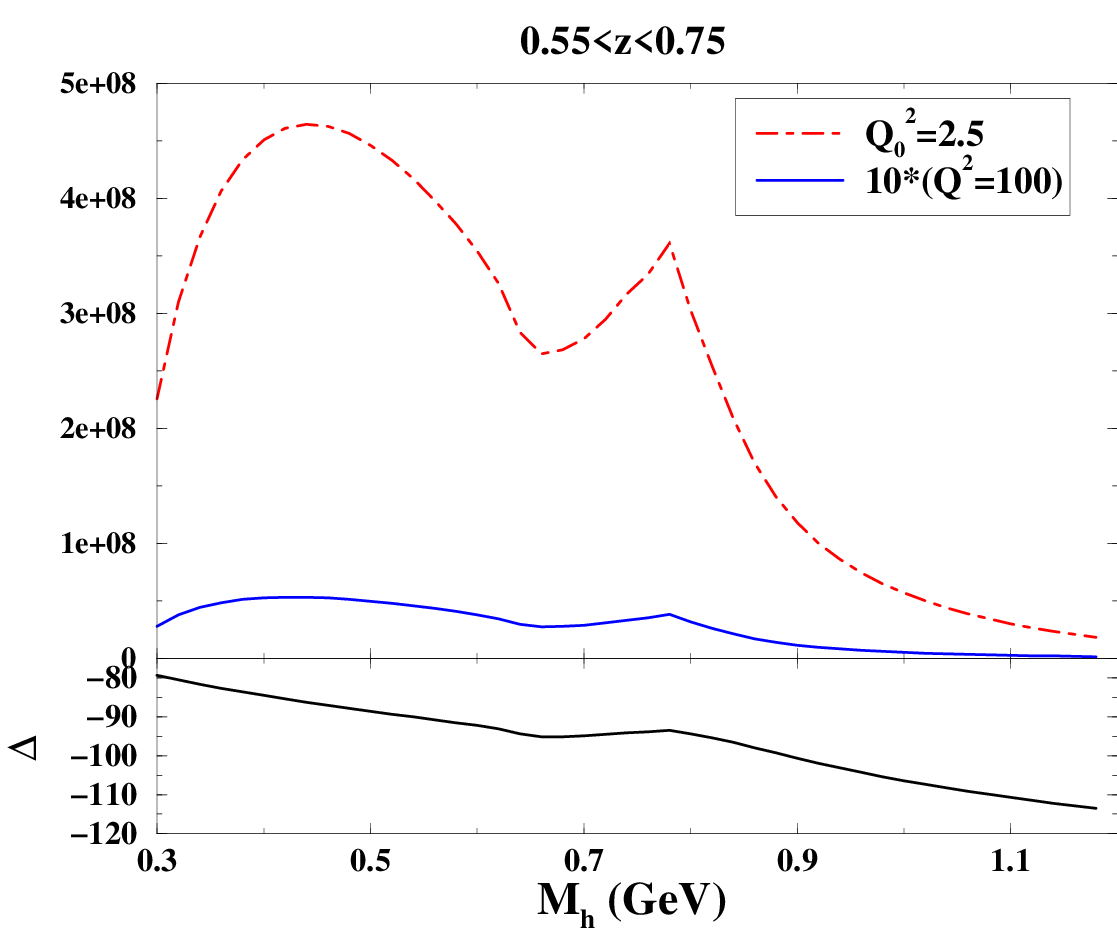}
\end{center}
\caption{Numerator (left panels) and denominator (right panels) of the 
azimuthal asymmetry in the same conditions and with the same notations as in 
Fig.~\protect{\ref{fig:AMhz}}, for the indicated boundary bins in $z$. The 
result at $Q^2=100$ GeV$^2$ (solid line in the upper plot of each panel) is 
emphasized by the factor 10.}
\label{fig:num-den}
\end{figure}

After the comments about Fig.~\ref{fig:dglap}, one would be tempted to 
attribute this sensitivity of the $M_h$ dependence to the hard scale $Q^2$ as 
coming from a spurious effect; indeed, in each panel of Fig.~\ref{fig:AMhz} the 
asymmetry is integrated in the indicated $z$ bin, which is obviously just a 
small fraction of the available phase space. However, the asymmetry of 
Eq.~(\ref{eq:asye+e-}) is the ratio of two objects that behave very differently 
under evolution because of their kernels $P(u)$ and $\delta P(u)$, 
respectively. Hence, there is no fundamental reason to expect the pure $M_h$ 
dependence of the asymmetry be preserved by DGLAP evolution, even after 
integrating upon the whole phase space of the other variables. Moreover, the 
moderate sensitivity of $A(z_{\mathrm{bin}},M_h^2)$ to the hard scale of the 
process is the result of a compensation of two very large sensitivities both in 
the numerator and in the denominator, as is clear from Fig.~\ref{fig:num-den}. 

%%%%%%%%%%%%%%%%%%%%%%%%%%%%%%%%% Fig. 5 %%%%%%%%%%%%%%%%%%%%%%%%%%%%%%%

%%%%%%%%%%%%%%%%%%%%%%%%%%%%%%%%%%%%%%%%%%%%%%%%%%%%%%%%%%%%%%%%%%%%%%%%%

In Fig.~\ref{fig:num-den}, the numerator (left panels) and denominator (right 
panels) of the asymmetry~(\ref{eq:asye+e-}) are shown with the same notations 
as in Fig.~\ref{fig:AMhz} for the $0.01\leq z \leq 0.1$ (upper panels) and 
$0.55 \leq z \leq 0.75$ bins (lower panels). The solid line, corresponding to 
the result at $Q^2=100$ GeV$^2$, is amplified by a factor 10. Therefore, the 
effect of DGLAP evolution is enormous, both in the numerator and in the 
denominator, where, in particular, it can reach a reduction factor of more than 
two orders of magnitude. Also the shape of the $M_h$ dependence is altered, 
making the more or less stable trend of $A(z_{\mathrm{bin}},M_h^2)$ at 
different $Q^2$ just a fortuitous case. 

In summary, even if DGLAP evolution of extended DiFF seems to mildly affect the 
predictions for the azimuthal asymmetry at BELLE, this small sensitivity arises 
from a dramatic compensation between big modifications in the numerator and in 
the denominator of the asymmetry. Therefore, it is wise to carefully consider 
such effect, because it could provide more sizeable modifications in other 
portions of the phase space. 

%%%%%%%%%%%%%%%%%%%%%%%%%%%%%%%%% Fig. 6 %%%%%%%%%%%%%%%%%%%%%%%%%%%%%%%

%%%%%%%%%%%%%%%%%%%%%%%%%%%%%%%%%%%%%%%%%%%%%%%%%%%%%%%%%%%%%%%%%%%%%%%%%

For sake of completeness, in Fig.~\ref{fig:Azzb} the azimuthal 
asymmetry~(\ref{eq:asye+e-}) is displayed as a function of $z$ for the 
$\overline{z}$ bins $[0.01,0.1]$, $[0.1,0.2]$, $[0.2,0.3]$, $[0.3,0.4]$, 
$[0.4,0.55]$, and $[0.55,0.75]$, with the same notations as in 
Fig.~\ref{fig:AMhz} and again after integrating upon 
$0.4\leq M_h, \overline{M}_h\leq 1.2$ GeV, and $-0.6\leq \cos\theta_2\leq 0.9$. 
It shows a rising trend for increasing both $z$ and $\overline{z}$. The effect 
of DGLAP evolution is small and within 10\%, except for the lowest $z$ values.

%%%%%%%%%%%%%%%%%%%%%%%%%%%%%%%%%%%%%%%%%%%%%%%%%%%%%%%%%%%%
\section{Conclusions}
\label{sec:conclusions}

In this paper, using the model calculation of extended dihadron fragmentation 
functions (DiFF) from Ref.~\cite{Bacchetta:2006un} we fitted the spin asymmetry 
recently extracted by the HERMES collaboration for the semi-inclusive 
deep-inelastic scattering (SIDIS) production of $(\pi^+ \pi^-)$ pairs on 
transversely polarized protons~\cite{Airapetian:2008sk}. Then, using the 
results of Ref.~\cite{Ceccopieri:2007ip} we calculated the evolution of 
extended DiFF at leading logarithm level, starting from the HERMES scale up 
to the scale of the process 
$e^+ e^- \rightarrow (\pi^+ \pi^-)_{\rm jet1}\, (\pi^+ \pi^-)_{\rm jet2} X$ at
BELLE kinematics. Finally, we made our predictions for the so-called 
Artru--Collins asymmetry, describing the correlation of angular distributions 
of the involved two pion pairs. The BELLE collaboration plans to measure this
asymmetry in the near future.

The absolute size of the Artru--Collins asymmetry turns out to be small, but
it should be within reach of the BELLE experimental capabilities, if compared 
with the corresponding Collins effect for two separated single-hadron 
fragmentations~\cite{Seidl:2008xc}. The theoretical error band, originating 
from the uncertainty in the fit of the SIDIS data, is always larger than the 
effects produced by the evolution of DiFF. The latter seems to mildly affect 
the predictions for the azimuthal asymmetry at BELLE. Nevertheless, this small 
sensitivity arises from a dramatic compensation between big modifications in the 
numerator and in the denominator of the asymmetry. Therefore, it is wise to 
carefully consider such effect, because it could provide more sizeable 
modifications in other portions of the phase space. 

We stress that azimuthal asymmetries based on the mechanism of dihadron 
fragmentation can be studied using collinear factorization, which facilitates 
the analysis of, e.g., evolution. From the theoretical point of view, this 
feature makes them a cleaner observable than the Collins effect in 
single-hadron fragmentation, where transverse-momentum dependent (TMD) 
functions are involved, whose evolution is yet not taken into account. All this 
procedure would not be plagued by theoretical uncertainties about factorization 
and evolution of TMD parton densities, which currently affect the analysis of 
single-hadron fragmentation. As a consequence, the option of using the 
semi-inclusive production of hadron pairs inside the same jet seems 
theoretically the cleanest way to extract the transversity distribution $h_1$, 
at present~\cite{Boer:2008mz}. 

%In fact, so far extended DiFF were modelled and fitted to SIDIS data using some 
%input for the transversity distribution, then evolved and used to make predictions for 
%the asymmetry in $e^+ e^-$ annihilations, as described above. But 
%when BELLE data will be
%available for the latter asymmetry, it will be possible to follow the reversed direction. 
When BELLE data will be available, it will be possible to constrain extended
DiFF on $e^+e^-$ data, to evolve them back to the HERMES scale, and to use them 
in the formula for the SIDIS spin asymmetry to directly extract $h_1$.

%%%%%%%%%%%%%%%%%%%%%%%%%%%%%%%%%%%%%%%%%%%%%%%%%%%%%%%%%%%%%%%%%%%%%%%%%%%%%%
\begin{acknowledgments}

This work is part of the European Integrated Infrastructure Initiative in
Hadronic Physics project under Contract No. RII3-CT-2004-506078.

Authored by Jefferson Science Associates, LLC under U.S. DOE Contract
No. DE-AC05-06OR23177. The U.S. Government retains a non-exclusive, paid-up,
irrevocable, world-wide license to publish or reproduce this manuscript for
U.S. Government purposes. 

A.M. acknowledges support from BRNS, government of India, and hospitality of 
INFN - Sezione di Pavia (Italy) and Jefferson Laboratory (Virginia - USA), 
where part of this work was done. 

\end{acknowledgments}
%%%%%%%%%%%%%%%%%%%%%%%%%%%%%%%%%%%%%%%%%%%%%%%%%%%%%%%%%%%%%%%%%%%%%%%%

%%%%%%%%%%%%%%%%%%%%%%%%%%%%%%%%%%%%%%%%%%%%%%%%%%%%%%%%%%%%%%%%%%%%%%%%%%%
\bibliographystyle{h-physrev}
\bibliography{mybiblio}

\begin{thebibliography}{10}

\bibitem{Konishi:1978yx}
K.~Konishi, A.~Ukawa, and G.~Veneziano,
\newblock Phys. Lett. {\bf B78}, 243 (1978).

\bibitem{deFlorian:2003cg}
D.~de~Florian and L.~Vanni,
\newblock Phys. Lett. {\bf B578}, 139 (2004), hep-ph/0310196.

\bibitem{Acton:1992sa}
OPAL, P.~D. Acton {\em et~al.},
\newblock Z. Phys. {\bf C56}, 521 (1992).

\bibitem{Abreu:1992xx}
DELPHI, P.~Abreu {\em et~al.},
\newblock Phys. Lett. {\bf B298}, 236 (1993).

\bibitem{Buskulic:1995gm}
ALEPH, D.~Buskulic {\em et~al.},
\newblock Z. Phys. {\bf C69}, 379 (1996).

\bibitem{Grazzini:1997ih}
M.~Grazzini, L.~Trentadue, and G.~Veneziano,
\newblock Nucl. Phys. {\bf B519}, 394 (1998), hep-ph/9709452.

\bibitem{Bianconi:1999cd}
A.~Bianconi, S.~Boffi, R.~Jakob, and M.~Radici,
\newblock Phys. Rev. {\bf D62}, 034008 (2000), hep-ph/9907475.

\bibitem{Bacchetta:2003vn}
A.~Bacchetta and M.~Radici,
\newblock Phys. Rev. {\bf D69}, 074026 (2004), hep-ph/0311173.

\bibitem{Ceccopieri:2007ip}
F.~A. Ceccopieri, M.~Radici, and A.~Bacchetta,
\newblock Phys. Lett. {\bf B650}, 81 (2007), hep-ph/0703265.

\bibitem{Efremov:1992pe}
A.~V. Efremov, L.~Mankiewicz, and N.~A. Tornqvist,
\newblock Phys. Lett. {\bf B284}, 394 (1992).

\bibitem{Collins:1994kq}
J.~C. Collins, S.~F. Heppelmann, and G.~A. Ladinsky,
\newblock Nucl. Phys. {\bf B420}, 565 (1994), hep-ph/9305309.

\bibitem{Artru:1996zu}
X.~Artru and J.~C. Collins,
\newblock Z. Phys. {\bf C69}, 277 (1996), hep-ph/9504220.

\bibitem{Barone:2003fy}
V.~Barone and P.~G. Ratcliffe,
\newblock {\em Transverse Spin Physics} (World Scientific, River Edge, USA,
  2003).

\bibitem{Ralston:1979ys}
J.~P. Ralston and D.~E. Soper,
\newblock Nucl. Phys. {\bf B152}, 109 (1979).

\bibitem{Martin:1999mg}
O.~Martin, A.~Schafer, M.~Stratmann, and W.~Vogelsang,
\newblock Phys. Rev. {\bf D60}, 117502 (1999), hep-ph/9902250.

\bibitem{Mukherjee:2003pf}
A.~Mukherjee, M.~Stratmann, and W.~Vogelsang,
\newblock Phys. Rev. {\bf D67}, 114006 (2003), hep-ph/0303226.

\bibitem{Mukherjee:2005rw}
A.~Mukherjee, M.~Stratmann, and W.~Vogelsang,
\newblock Phys. Rev. {\bf D72}, 034011 (2005), hep-ph/0506315.

\bibitem{Collins:1993kk}
J.~C. Collins,
\newblock Nucl. Phys. {\bf B396}, 161 (1993), hep-ph/9208213.

\bibitem{Airapetian:2004tw}
HERMES, A.~Airapetian {\em et~al.},
\newblock Phys. Rev. Lett. {\bf 94}, 012002 (2005), hep-ex/0408013.

\bibitem{Diefenthaler:2007}
HERMES, M.~Diefenthaler,
\newblock (2007), arXiv:0706.2242 [hep-ex],
\newblock Proceedings of the 15th International Workshop on Deep Inelastic
  Scattering (DIS 2007), Munich, Germany, 16 - 20 Apr 2007.

\bibitem{Alexakhin:2005iw}
COMPASS, V.~Y. Alexakhin {\em et~al.},
\newblock Phys. Rev. Lett. {\bf 94}, 202002 (2005), hep-ex/0503002.

\bibitem{Ageev:2006da}
COMPASS, E.~S. Ageev {\em et~al.},
\newblock Nucl. Phys. {\bf B765}, 31 (2007), hep-ex/0610068.

\bibitem{Levorato:2008tv}
COMPASS, S.~Levorato,
\newblock (2008), 0808.0086.

\bibitem{Boer:1997mf}
D.~Boer, R.~Jakob, and P.~J. Mulders,
\newblock Nucl. Phys. {\bf B504}, 345 (1997), hep-ph/9702281.

\bibitem{Seidl:2008xc}
Belle, R.~Seidl {\em et~al.},
\newblock Phys. Rev. {\bf D78}, 032011 (2008), 0805.2975.

\bibitem{Anselmino:2007fs}
M.~Anselmino {\em et~al.},
\newblock Phys. Rev. {\bf D75}, 054032 (2007), hep-ph/0701006.

\bibitem{Boer:1997nt}
D.~Boer and P.~J. Mulders,
\newblock Phys. Rev. {\bf D57}, 5780 (1998), hep-ph/9711485.

\bibitem{Boer:2008fr}
D.~Boer,
\newblock (2008), 0804.2408.

\bibitem{Collins:1981uk}
J.~C. Collins and D.~E. Soper,
\newblock Nucl. Phys. {\bf B193}, 381 (1981).

\bibitem{Ji:2004wu}
X.~Ji, J.-P. Ma, and F.~Yuan,
\newblock Phys. Rev. {\bf D71}, 034005 (2005), hep-ph/0404183.

\bibitem{Ceccopieri:2005zz}
F.~A. Ceccopieri and L.~Trentadue,
\newblock Phys. Lett. {\bf B636}, 310 (2006), hep-ph/0512372.

\bibitem{Bacchetta:2007wc}
A.~Bacchetta, L.~P. Gamberg, G.~R. Goldstein, and A.~Mukherjee,
\newblock Phys. Lett. {\bf B659}, 234 (2008), 0707.3372.

\bibitem{Boer:2001he}
D.~Boer,
\newblock Nucl. Phys. {\bf B603}, 195 (2001), hep-ph/0102071.

\bibitem{Boer:2008mz}
D.~Boer,
\newblock (2008), 0808.2886.

\bibitem{Collins:1994ax}
J.~C. Collins and G.~A. Ladinsky,
\newblock (1994), hep-ph/9411444.

\bibitem{Jaffe:1998hf}
R.~L. Jaffe, X.~Jin, and J.~Tang,
\newblock Phys. Rev. Lett. {\bf 80}, 1166 (1998), hep-ph/9709322.

\bibitem{Radici:2001na}
M.~Radici, R.~Jakob, and A.~Bianconi,
\newblock Phys. Rev. {\bf D65}, 074031 (2002), hep-ph/0110252.

\bibitem{Bacchetta:2006un}
A.~Bacchetta and M.~Radici,
\newblock Phys. Rev. {\bf D74}, 114007 (2006), hep-ph/0608037.

\bibitem{Bacchetta:2002ux}
A.~Bacchetta and M.~Radici,
\newblock Phys. Rev. {\bf D67}, 094002 (2003), hep-ph/0212300.

\bibitem{Airapetian:2008sk}
HERMES, A.~Airapetian {\em et~al.},
\newblock JHEP {\bf 06}, 017 (2008), 0803.2367.

\bibitem{Martin:2007au}
COMPASS, A.~Martin,
\newblock (2007), hep-ex/0702002.

\bibitem{Boer:2003ya}
D.~Boer, R.~Jakob, and M.~Radici,
\newblock Phys. Rev. {\bf D67}, 094003 (2003), hep-ph/0302232.

\bibitem{Abe:2005zx}
BELLE, K.~Abe {\em et~al.},
\newblock Phys. Rev. Lett. {\bf 96}, 232002 (2006), hep-ex/0507063.

\bibitem{Hasuko:2003ay}
K.~Hasuko, M.~Grosse~Perdekamp, A.~Ogawa, J.~S. Lange, and V.~Siegle,
\newblock AIP Conf. Proc. {\bf 675}, 454 (2003).

\bibitem{Bacchetta:2004it}
A.~Bacchetta and M.~Radici,
\newblock Phys. Rev. {\bf D70}, 094032 (2004), hep-ph/0409174.

\bibitem{Yang:2008}
STAR, R.~Yang,
\newblock (2008),
\newblock PKU-RBRC Workshop on Transverse Spin Physics, Beijing, June 30th-July
  4th, 2008, http://rchep.pku.edu.cn/workshop/0806/20080627-iff.pdf.

\bibitem{Bacchetta:2004jz}
A.~Bacchetta, U.~D'Alesio, M.~Diehl, and C.~A. Miller,
\newblock Phys. Rev. {\bf D70}, 117504 (2004), hep-ph/0410050.

\bibitem{Anselmino:2008sj}
M.~Anselmino {\em et~al.},
\newblock (2008), 0807.0173.

\bibitem{Altarelli:1977zs}
G.~Altarelli and G.~Parisi,
\newblock Nucl. Phys. {\bf B126}, 298 (1977).

\bibitem{Stratmann:2001pt}
M.~Stratmann and W.~Vogelsang,
\newblock Phys. Rev. {\bf D65}, 057502 (2002), hep-ph/0108241.

\bibitem{Artru:1990zv}
X.~Artru and M.~Mekhfi,
\newblock Z. Phys. {\bf C45}, 669 (1990).

\end{thebibliography}
%%%%%%%%%%%%%%%%%%%%%%%%%%%%%%%%%%%%%%%%%%%%%%%%%%%%%%%%%%%%%%%%%%%%%%%%%%%%%

\end{document}